\documentclass[preprint,aps,showpacs,nofootinbib,preprintnumbers,amsmath,amssymb]{revtex4-1}
\usepackage{amssymb}
\usepackage{epsfig}
\usepackage{graphicx}
\usepackage{subfigure}
\usepackage{dcolumn}
\usepackage{bm}
\usepackage[usenames ,dvipsnames]{xcolor}
\usepackage{slashed}

\begin{document}

\title{Spontaneous $CP$-Violating Electroweak Baryogenesis and Dark Matter from a Complex Singlet Scalar}

\author{Bohdan Grzadkowski$^{1}$\footnote{bohdan.grzadkowski@fuw.edu.pl}
}
 \affiliation{$^{1}$Institute of Theoretical Physics, Faculty of Physics, University of Warsaw, Pasteura 5, 02-093 Warsaw, Poland}

\author{Da~Huang$^{1}$\footnote{dahuang@fuw.edu.pl}
}
 \affiliation{$^{1}$Institute of Theoretical Physics, Faculty of Physics, University of Warsaw, Pasteura 5, 02-093 Warsaw, Poland}


\date{\today}
\begin{abstract}
CP non-invariance is strongly limited by present experiments, while extra sources of CP-violation are needed for a successful baryogenesis.
Motivated by those observations we consider a model which predicts spontaneous violation of CP at high temperature and restoration of CP
at present temperature of the Universe. In addition we propose a dark matter (DM) candidate that meets all known properties of DM. 
Looking for a minimal model that satisfies the above conditions leads us to extending the Standard Model (SM) of fundamental interactions by adding 
a complex singlet scalar $S$. We impose the $CP$ and $Z_2$ symmetries on the scalar potential. 
With the complex vacuum expectation value of $S$ at the temperature higher than the EW phase transition, the $CP$ symmetry is spontaneously broken and a strong first-order electro-weak phase transition is easily realized. Introducing a dimension-6 effective operator that gives new complex contributions to the top quark mass, we show that it is easy to yield the observed baryon asymmetry in our Universe. On the other hand,  the $CP$ and $Z_2$ symmetries are recovered after the EW phase transition 
so that the present strong constraints on $CP$ violation can be satisfied and
the lighter of $\Re S$ or $\Im S$ can be the dark matter candidate. 
By scanning the parameter space, we find regions where the model can explain the dark matter relic abundance and the baryon asymmetry simultaneously while satisfying all other experimental constraints. Finally, we discuss the explicit $CP$ symmetry breaking in the scalar potential that can help dynamically eliminate the domains producing the negative baryon asymmetry. It is found that this can be achieved by a tiny explicit $CP$-violating phase of ${\cal O}(10^{-15})$.   
\end{abstract}

\maketitle

\section{Introduction}
\label{s1}
In spite of the great success of the Standard Model (SM) of particle physics in explaining the present visible Universe, there are still many puzzles awaiting to be understood. Among them, two prominent mysteries are the observation of dark matter (DM)~\cite{pdg,Bergstrom:2012fi} and the origin of matter-antimatter asymmetry in the Universe~\cite{Ade:2015xua}. On one hand, the existence of DM has been firmly established by measurements of galaxy rotation curves~\cite{Rubin:1970zza}, gravitational lensing effects~\cite{Jee:2007nx}, and cosmic microwave background (CMB)~\cite{Ade:2015xua}. 
However, the nature of DM is still out of reach. On the other hand, the observed baryon asymmetry is usually represented in terms of the following baryon-to-entropy ratio~\cite{Ade:2015xua}
\begin{eqnarray}
\eta_B \equiv \frac{n_B}{s} = (8.61\pm 0.09)\times 10^{-11}\,,
\end{eqnarray}
where $n_B$ and $s$ are the densities of baryon number and entropy of the Universe. It is well-known that a successful baryogeneis theory should satisfy the three Sakharov criteria~\cite{Sakharov:1967dj}: (1) baryon number violation; (2) $C$ and $CP$ violations; and (3) a departure from the thermodynamic equilibrium. One intriguing mechanism is provided by the electroweak (EW) baryogenesis~\cite{Kuzmin:1985mm,Cohen:1990py,Cohen:1993nk,Rubakov:1996vz, Cline:2000fh,Morrissey:2012db,Konstandin:2013caa}. In this framework, when the strong first-order EW phase transition (EWPT) occurs, the baryon-number-violating EW sphaleron processes~\cite{Klinkhamer:1984di,Gavela:1994dt,Huet:1994jb} can bias the $CP$ asymmetry produced around the EWPT bubble wall into the baryon asymmetry. Unfortunately, in the SM, the EWPT is found to be a crossover~\cite{Kajantie:1996mn,Csikor:1998eu,Aoki:1999fi}, and the $CP$ violation provided by the CKM matrix is too small to account for the observed asymmetry~\cite{Shaposhnikov:1987tw, Farrar:1993hn,Gavela:1993ts,Konstandin:2003dx}. Therefore, both the DM and EW baryogensis require new physics beyond the SM.

In the present work, we try to explain the observed DM relic density and the baryon asymmetry simultaneously by extending the SM by a complex EW singlet scalar $S$~\cite{McDonald:1993ey,McDonald:1995hp,Barger:2008jx,Profumo:2007wc,Branco:1998yk, Gonderinger:2012rd,Coimbra:2013qq,Costa:2014qga,Jiang:2015cwa,Chao:2017oux,Chiang:2017nmu}. Note that one condition for the EW baryogenesis is that the first-order EWPT should be strong enough, which is usually parametrized by~\cite{Moore:1998swa}
\begin{eqnarray}\label{xi}
 \frac{v_c}{T_c} > 1\,,
\end{eqnarray}  
where $v_c$ is the SM Higgs vacuum expectation value (VEV) at the critical temperature $T_c$. This condition guarantees the produced baryon asymmetry in the EW symmetry-breaking phase is not washed out by the EW sphalerons. Recently reliability and gauge invariance of Eq.~(\ref{xi}) have been questioned in the literature~\cite{Patel:2011th}. Moreover, additional $CP$ violating (CPV) interactions required by the baryogenesis are severely constrained by the negative results in the electric dipole moment (EDM) searches for electrons~\cite{Baron:2013eja} and neutrons~\cite{pdg}. Both conditions can be easily satisfied by introducing an extra complex scalar singlet $S$ and imposing the $Z_2$ and $CP$ symmetries. In our model, the phase transition (PT) follows a two-step pattern in which $S$ firstly acquires a nonzero complex-valued VEV, and then the EWPT goes from $(0, w_c e^{i\alpha}/\sqrt{2})$ to $(v_c,0)$ where the two values in the bracket represent the VEVs of the SM Higgs and the singlet $(\langle h \rangle, \langle S \rangle)$. It will be shown that for the complex scalar, like in the case of its real singlet cousin~\cite{Espinosa:1993bs,Choi:1993cv,Ham:2004cf,Espinosa:2007qk,Ahriche:2007jp, Espinosa:2011ax,  Ahriche:2012ei, Profumo:2014opa, Alanne:2014bra, Alanne:2016wtx, Tenkanen:2016idg, Vaskonen:2016yiu, Cline:2012hg, Espinosa:2011eu}, there is a large barrier at tree level so that the EWPT can easily satisfy Eq.~(\ref{xi}). Furthermore, the complex $\langle S \rangle$, together with the following dimension-6 effective operator
\begin{eqnarray}\label{O6}
{\cal O}_6 = \frac{S^2}{\Lambda^2} \bar{Q}_{3L} \tilde{H} t_R +{\rm H.c.}\,,
\end{eqnarray} 
breaks the $CP$ symmetry spontaneously, which is a necessary condition for the EW baryogenesis. Here $Q_{3L}$ and $t_R$ denote the third-generation left-handed quark doublet and right-handed top quark fields, $\tilde{H} = i\sigma_2 H^*$, and $\Lambda$ is a cutoff scale parametrizing the amplitude of this effective operator. After the EWPT, the $Z_2$ and $CP$ symmetries are restored, so that the lighter real component of $S$ can be an ideal DM candidate stabilized by the $Z_2$ symmetry, and strong constraints for $CP$ violations are naturally avoided~\cite{McDonald:1993ey, Chao:2017oux}. This model can be regarded as a realization of the finite-temperature spontaneous CPV EW baryogenesis mechanism proposed in Refs.~\cite{McDonald:1993ey, Chao:2017oux,Comelli:1993ne}. Also, we would like to mention that top-related effective operators similar to ${\cal O}_6$ for real and complex singlet scalars have already been discussed in Refs.~\cite{Vaskonen:2016yiu,Cline:2012hg,Espinosa:2011eu,Jiang:2015cwa}.

The paper is organized as follows. In Sec~\ref{Model}, we present our model and analyze its strong first-order EWPT. Then we discuss the DM phenomenology and EW baryogenesis in the Sec.~\ref{DMPheno} and \ref{BAU}, where a large-scale random scan of parameter space is performed. Unfortunately, we were unable to find parameters which could accommodate the DM relic abundance and EW baryogenesis simultaneously. In order to search for such models, we perform a random scan again in Sec.~\ref{GoodModel}, by focusing on the region where DM particles annihilate mainly through the SM Higgs resonance. One problem that always plagued  models with spontaneous CPV baryogenesis is the appearance of the domains of EW symmetry-breaking vacua which give rise to the antibaryon number excesses during the EW baryogenesis~\cite{McDonald:1993ey,McDonald:1995hp, Chao:2017oux,Comelli:1993ne}. In Sec.~\ref{dw}, we show how to eliminate these domains by introducing a very tiny explicit CPV phase in the scalar potential. Finally, we conclude in Sec.~\ref{conclusion}.    

\section{The Model and Electroweak Phase Transition}\label{Model}
The model extends the SM by addition of a complex scalar $S = (s+ia)/\sqrt{2}$ that is odd under a $Z_2$ symmetry in order to guarantee the stability of the lighter of $S$ components. We also assume the $CP$ symmetry in the dark sector so that the couplings involving $S$ should be real. Thus, the extended scalar potential at zero temperature can be written as follows:
\begin{eqnarray}\label{V0}
V_0 (H,S) &=& {\lambda_H} \left(|H|^2 - \frac{v_0^2}{2}\right)^2 -\mu_1^2 (S^* S)^2 - \frac{\mu^2_2}{2} (S^2+S^{*2})\nonumber\\
&& +{\lambda_1}(S^* S)^2 + \frac{\lambda_2}{4}(S^2+S^{*2})^2  + \frac{\lambda_3}{2} |S|^2 (S^2 + S^{*2})\nonumber\\
&& +|H|^2 \left[ \kappa_1 (S^* S) + \frac{\kappa_2}{2}(S^2 + S^{*2}) \right]  \nonumber\\
&=& -\frac{1}{2} \lambda_H v_0^2 h^2 + \frac{1}{4}\lambda_H h^4 -\frac{1}{2} (\mu_1^2+\mu_2^2) s^2 - \frac{1}{2}(\mu_1^2 - \mu_2^2)a^2  \nonumber\\
&& + \frac{1}{4}\left({\lambda_1} + \lambda_2 + \lambda_3\right) s^4 + \frac{1}{4}\left({\lambda_1} + \lambda_2 - \lambda_3\right) a^4 \nonumber\\
&& + \frac{1}{4} (\kappa_1 + \kappa_2) h^2 s^2 + \frac{1}{4} (\kappa_1 - \kappa_2) h^2 a^2 +\frac{1}{2}( \lambda_1 - \lambda_2) s^2 a^2 + {\rm const.} \,.
\end{eqnarray}
where $H = (0,h/\sqrt{2})^T$ represents the SM Higgs doublet written in the unitary gauge. In order for the later convenience, we have expanded the Lagrangian in terms of the components $h$, $s$ and $a$. It is easy to see that the final potential is a function of $h^2$, $s^2$ and $a^2$, which can be traced back to the assigned $Z_2$ and $CP$ symmetries. Since we are interested in the PT in this model, we need to calculate the leading-order finite-temperature corrections in the high-temperature expansion, which is given by
\begin{eqnarray}
V_T = \frac{1}{2}c_h T^2 h^2 + \frac{1}{2} c_s T^2 s^2 + \frac{1}{2} c_a T^2 a^2\,,
\end{eqnarray}
where
\begin{eqnarray}
c_h &=& \frac{3g^2}{16}  + \frac{g^{\prime 2}}{16} + \frac{y_t^2}{4} + \frac{\lambda_H}{2} + \frac{\kappa_1}{12}\,,\nonumber\\
c_s &=& \frac{1}{6}(2\lambda_1 + \kappa_1 + \kappa_2)+\frac{\lambda_3}{4}\,,\nonumber\\
c_a &=& \frac{1}{6}(2\lambda_1 + \kappa_1 - \kappa_2)-\frac{\lambda_3}{4}\,.
\end{eqnarray}
Altogether, the total finite-temperature Lagrangian is $V_{\rm tot} = V_0 + V_T$.

It has recently been pointed out in Ref.~\cite{Haber:2012np} that one necessary condition for a theory with a complex scalar $S$ to achieve spontaneous $CP$ violation is the $U(1)$ symmetry related to $S$ is explicitly broken in the scalar potential by at least two terms different dimesnion. It is obvious that the Lagrangian in Eq.~(\ref{V0}) satisfies this condition, which may break $CP$ symmetry by the complex VEV of $S$. In the present paper, we explore the possible EWPT from a CPV EW-symmetric vacuum with $(0, w_c e^{i\alpha}/\sqrt{2})$ to the $CP$-symmetric EW-broken vacuum $(v_c, 0)$, in which the two entries in the bracket denote the VEVs of the SM Higgs $\langle h \rangle$ and the singlet $\langle S \rangle$. In order to describe this PT, we follow the method in Ref.~\cite{Espinosa:2011eu} by rewriting the finite-temperature potential as follows: 
\begin{eqnarray}\label{V_tot}
V_{\rm tot} &=& \frac{\lambda_{hs}}{4} \left(h^2-v_c^2 + \frac{v_c^2 s^2}{w_c^2 \cos^2 \alpha} \right)^2 + \frac{\lambda_{ha}}{4}\left(h^2-v_c^2 + \frac{v_c^2 a^2}{w_c^2 \sin^2 \alpha} \right)^2\nonumber\\
&& + \frac{\lambda_{sa}}{4}\left( s^2 \sin^2\alpha -a^2 \cos^2\alpha \right)^2 +\frac{\kappa_{hs}}{4} h^2 s^2 + \frac{\kappa_{ha}}{4} h^2 a^2 \nonumber\\
&& + \frac{1}{2} (T^2 - T_c^2) [c_h h^2 + c_s s^2 + c_a a^2] \nonumber\\
&=& -\frac{1}{2} \left[{(\lambda_{hs} + \lambda_{ha})} v_c^2 + c_h T_c^2\right]h^2 - \frac{1}{2} \left( \frac{\lambda_{hs} v_c^4}{w_c^2 \cos^2\alpha} + c_s T_c^2 \right) s^2 \nonumber\\
&& -\frac{1}{2} \left( \frac{\lambda_{ha} v_c^4}{w_c^2 \sin^2\alpha} + c_a T_c^2 \right) a^2  + \frac{T^2}{2}(c_h h^2 + c_s s^2 + c_a a^2) \nonumber\\
&& + \frac{1}{4} (\lambda_{hs}+\lambda_{ha})h^4 + \frac{1}{4}\left(\lambda_{sa} \sin^4 \alpha + \frac{\lambda_{hs} v_c^4}{w_c^4 \cos^4\alpha}\right) s^4 + \frac{1}{4}\left({\lambda_{sa}\cos^4\alpha} + \frac{\lambda_{ha} v_c^4}{w_c^4 \sin^4\alpha}\right) a^4 \nonumber\\
&& + \frac{1}{4} \left(\kappa_{hs}+\frac{2 \lambda_{hs} v_c^2}{w_c^2 \cos^2\alpha}\right)h^2 s^2
 + \frac{1}{4} \left(\kappa_{ha}+\frac{2 \lambda_{ha} v_c^2}{w_c^2 \sin^2\alpha}\right)h^2 a^2 - \frac{\lambda_{sa}}{2}  \sin^2\alpha \cos^2\alpha s^2 a^2\,.
\end{eqnarray}
By comparing the second lines in Eqs.~(\ref{V0}) and (\ref{V_tot}) at $T=0$, we can read off the critical temperature for the first-order EWPT
\begin{eqnarray}\label{Tc}
T_c^2 = \lambda_H(v_0^2 - v_c^2)/c_h\,,
\end{eqnarray}
and the following relations among various parameters
\begin{eqnarray}\label{LamKap}
&&\lambda_H = \lambda_{hs} + \lambda_{ha}\,, \nonumber \\
&& \kappa_1 = \frac{1}{2}(\kappa_{hs} + \kappa_{ha}) + \frac{v_c^2}{w_c^2} \left(\frac{\lambda_{hs}}{\cos^2 \alpha} + \frac{\lambda_{ha}}{\sin^2\alpha} \right)\,,\, 
\kappa_2 = \frac{1}{2}(\kappa_{hs} - \kappa_{ha}) + \frac{v_c^2}{w_c^2} \left(\frac{\lambda_{hs}}{\cos^2 \alpha} - \frac{\lambda_{ha}}{\sin^2\alpha} \right)\,, \nonumber\\
&& \lambda_1 = \frac{\lambda_{sa} \cos^2 (2\alpha)}{4} + \frac{v_c^4}{4 w_c^4}\left( \frac{\lambda_{hs}}{\cos^4\alpha} + \frac{\lambda_{ha}}{\sin^4\alpha} \right) \,, \, 
\lambda_2 = \frac{\lambda_{sa}}{4} + \frac{v_c^4}{4 w_c^4}\left( \frac{\lambda_{hs}}{\cos^4\alpha} + \frac{\lambda_{ha}}{\sin^4\alpha} \right) \,, \nonumber\\
&& \lambda_3 = \frac{\lambda_{sa}}{2}(\sin^4\alpha - \cos^4 \alpha) + \frac{v_c^4}{2 w_c^4}\left(\frac{\lambda_{hs}}{\cos^4\alpha} - \frac{\lambda_{ha}}{\sin^4 \alpha}\right)\,.
\end{eqnarray}
We can also obtain the zero-temperature masses for the three scalars, $h$, $s$ and $a$, as follows
\begin{eqnarray}\label{Mass}
m_h^2 &=& 2\lambda_H v_0^2\,,\nonumber\\
m_s^2 &=& \frac{1}{2}\kappa_{hs} v_0^2 + \lambda_H (v_0^2 - v_c^2) \left(\frac{\lambda_{hs}v_c^2}{\lambda_H w_c^2 \cos^2 \alpha} - \frac{c_s}{c_h}\right)\,,\nonumber\\
m_a^2 &=& \frac{1}{2}\kappa_{ha} v_0^2 + \lambda_H (v_0^2 - v_c^2) \left(\frac{\lambda_{ha}v_c^2}{\lambda_H w_c^2 \sin^2 \alpha} - \frac{c_a}{c_h}\right)\,.
\end{eqnarray} 

The advantage to introduce the critical-temperature Lagrangian in Eq.~(\ref{V_tot}) is that it makes easier the analysis of the first-order EWPT. Here we assume that all the dimensionless couplings in Eq.~(\ref{V_tot}) are positive, thus the potential are absolutely stable at $T_c$ with $(0, w_c e^{i\alpha}/\sqrt{2})$ and $(v_c,0)$ the two vacua in the potential. Furthermore, the correct direction of the EWPT requires
\begin{eqnarray}
c_h v_c^2 > c_s w_c^2 \cos^2\alpha + c_a w_c^2 \sin^2\alpha\,. 
\end{eqnarray}
A further condition in Eq.~(\ref{xi}) is needed to ensure the EWPT strong enough in order to suppress baryon number washout effects in the EW broken phase. 

For simplicity, in our numerical scanning of parameter space, we use following 7 parameters in the Lagrangian of Eq.~(\ref{V_tot}) as free parameters
\begin{eqnarray}
\frac{v_0}{v_c}\,, \frac{v_c}{w_c}\,, \alpha\,, \lambda_{ha}\,, \lambda_{sa}\,,\kappa_{ha}\,,\kappa_{hs}\,,
\end{eqnarray}
while other parameters can be derived with relations in Eqs.~(\ref{Tc}-\ref{Mass}).
It is seen from Eq.~(\ref{Tc}) that the critical temperature $T_c$ exists as long as $v_0/v_c > 1$. Also, due to the $Z_2$ invariance of the effective operator ${\cal O}_6$, the two vacua with $\langle S \rangle = w_c e^{i\alpha}$ and $-\langle S\rangle = w_c e^{i(\alpha+\pi)}$ lead to the same CPV and thus the same baryon asymmetry. Therefore, without loss of generality, we can restrict the CPV phase $\alpha$ in the range of $[-\pi/2,\pi/2)$. Finally, note that the Lagrangian $V_0$ in Eq.~(\ref{V0}) is very useful for our discussion of particle phenomenology at zero temperature. In order to keep the perturbativity of the model at the EW scale, the dimensionless parameters in Eq.~(\ref{V0}) cannot be too large. For the purpose of illustration, we enforce these parameters to be $|\lambda_{1,2,3}, \kappa_{1,2}|\leqslant 5$~\cite{Nebot:2007bc}. 

\section{Dark Matter Phenomenology}\label{DMPheno}
After the EWPT, the $Z_2$ symmetry is recovered, so that the lightest $Z_2$-odd particle can be the DM candidate. In the present model, we denote the DM particle as $X$ which is the lighter scalar of $s$ and $a$. Note that the dark sector couples to the SM sector only through the interactions in the scalar potential, apart from the effective operator ${\cal O}_6$, so that the DM phenomenology is mainly determined by its coupling to $h$. In terms of parameters in Eq.~(\ref{V_tot}), we can rewrite this coupling to be $\lambda_{hX} h^2 X^2/4$ with
\begin{eqnarray}\label{Hportal}
\lambda_{hX} = \left\{\begin{array}{cc}
\kappa_{hs} + \frac{2\lambda_{hs}v_c^2}{w_c^2\cos^2\alpha}\,, & X=s\\
\kappa_{ha} + \frac{2\lambda_{ha}v_c^2}{w_c^2\sin^2\alpha}\,, & X=a
\end{array}\right..
\end{eqnarray}
After the spontaneous EW symmetry breaking, the above coupling can generate the triple-scalar interaction $(\lambda_{hX} v_0) h S^2$. 

With the above couplings, the DM $X$ relic abundance can be obtained via its annihilations into various SM particles by the SM Higgs exchange. In our numerical analysis, we apply the code {\tt MicrOMEGAs}~\cite{Belanger:2001fz,Belanger:2013oya} to perform such calculations. In this and next sections, we do not require the model to explain all of the DM relic density. Rather, we allow the DM to be subdominant which is parametrized by the following DM density fraction~\cite{Cline:2012hg}
\begin{eqnarray}
f_X = \frac{\Omega_{X}h^2}{\Omega_{\rm DM, obs} h^2}\,,
\end{eqnarray}
in which $\Omega_{\rm DM, obs} h^2 = 0.1186$ is the central value of the most recent DM abundance measurement by the Planck Collaboration~\cite{Ade:2015xua}.

The Higgs portal coupling in Eq.~(\ref{Hportal}) also induces the signals of DM direct and indirect detections. For the DM direct detection, the SM Higgs mediation gives the following spin-independent DM-nucleon ($XN$) scattering cross section
\begin{equation}
\sigma_{XN} = \frac{\lambda_{hX}^2 f_N^2}{4\pi} \frac{\mu^2_{XN} m_N^2}{m_X^2 m_h^4}\,,
\end{equation}
where $f_N = 0.3$ denotes the Higgs-nucleon coupling~\cite{Cline:2013gha, Alarcon:2011zs, Alarcon:2012nr}, and $\mu_{XN} = m_N m_X/(m_N+m_X)$ is the DM-nucleon reduced mass with $m_N$ being the nucleon mass. In the parameter space of interest, the latest XENON1T experiments~\cite{Aprile:2018dbl} set the most stringent constraint up to now. In order to directly compare with the experimental upper bounds, we would like to define the following effective DM-nucleon cross section~\cite{Cline:2012hg}
\begin{eqnarray}
\sigma_{XN}^{\rm eff} \equiv f_X \sigma_{XN}\,,
\end{eqnarray}
in order to take into account the situation when $X$ is subdominant as the DM relic density.

For the DM indirect detections, the DM annihilations through the Higgs portal also give rise to the $\gamma$-ray excesses in the spheroidal dwarf galaxies, $e^\pm$ signals in our galaxy, and modification of the ionization history of our Universe, which are strongly constrained by the observations from Fermi-LAT~\cite{Ackermann:2015zua}, AMS-02~\cite{AMS2t,AMS2ep} and Planck~\cite{Ade:2015xua} satellites. It is seen in Ref.~\cite{Elor:2015bho} that for DM mass above 1~GeV, the Fermi-LAT measurements of $\gamma$-rays from spheroidal dwarfs gives the strongest upper bound on the DM annihilations.  Moreover, note that the final products in the DM annihilations via the Higgs portal consists of $b\bar{b}$, $ZZ$, $W^+ W^-$, and light quark pairs. It is shown in Ref.~\cite{Ackermann:2015zua} that all of these channels yield almost the universal upper bounds for the DM annihilations. Therefore, we apply the Fermi-LAT constraints~\cite{Ackermann:2015zua} on the $b\bar{b}$ final state when $m_X \geqslant m_b$, while those on light quarks for the case with $m_X < m_b$.

Further constraints on our DM model are provided by collider searches. In particular, when $m_X < m_h/2$, the DM particle would lead to the invisible decay of the SM Higgs boson. The predicted Higgs invisible width is 
\begin{eqnarray}
\Gamma(h\to XX) = \frac{\lambda_{hX}^2 v_0^2}{32\pi m_h} \sqrt{1-\frac{4 m_X^2}{m_h^2}}\,,
\end{eqnarray} 
which should be compared with the current upper bound ${\rm Br}(h\to XX) \leqslant 0.24$~\cite{pdg}. Moreover, the CMS monojet search~\cite{Khachatryan:2014rra} can provide another test of the present model. We use the code incorporated in {\tt MicrOMEGAs}~\cite{Barducci:2016pcb} to exclude the parameter points at the $95\%$ C.L. with the CLs method~\cite{CLs1, CLs2}. 

\begin{figure}[]
\includegraphics[scale=0.48]{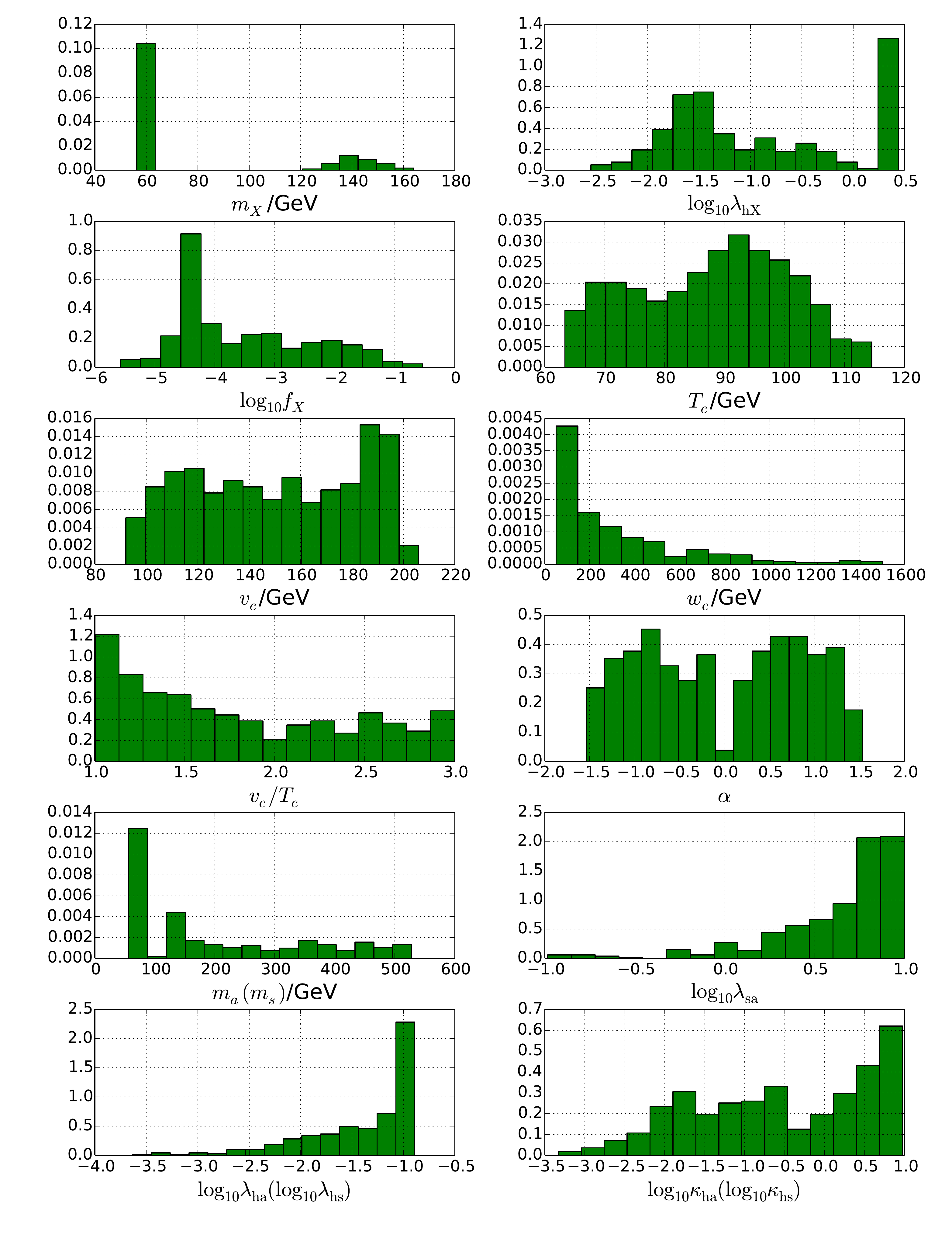}
\caption{The distributions of the parameters for the points satisfying the strongly first-order PT conditions and all the DM constraints. Note that since $s$ and $a$ are equivalent in the scalar potential, the distributions for the couplings $\lambda_{hs}$ and $\lambda_{ha}$ are essentially the same, so are for couplings $\kappa_{hs}$ and $\kappa_{ha}$. Thus, we only show one plot for either pair of couplings.}\label{dist}
\end{figure}

In our numerical study, we apply the random scan over the whole parameter space by taking into account all of the above constraints from the DM physics and the strong first-order PT. After a random scan over $2\times 10^8$ model parameter points where the input parameters vary in the following ranges:
\begin{eqnarray}
& &v_0/v_c = 1.0 \sim 10.0\,,\quad v_c/w_c = 0.1 \sim 10.0, \quad \alpha = -\pi/2 \sim \pi/2\,,\nonumber\\
&& \lambda_{ha} = 0 \sim \lambda_H\,,\quad \lambda_{sa} = 0 \sim 10.0\,,\quad \kappa_{hs,ha} = 0 \sim 10.0\,,
\end{eqnarray}
we find 388 points consistent with the DM searches and the strong first-order PT requirement. The distributions of the surviving points for various physical parameters of interest are shown in Fig.~\ref{dist}. One observes that the distribution of DM masses can be divided into two regions: (I) $55 \sim 65$~GeV, and (II) $120 \sim 170$~GeV. The DM-Higgs coupling are distributed in the range $10^{-3} < \lambda_{hX} < 3$ with the largest peak around 2 and a relatively small one at $0.03$. We find that all models can only give a subdominant DM relic density, with its typical fraction around $10^{-4} \sim 10^{-5}$. The values of $v_c$ lie in the range $90 \sim 210$~GeV with the peak at 190~GeV, while the  critical temperature $T_c$ is found to be between 63 GeV and 115 GeV. The parameter $v_c/T_c$ signalling the strength of the EWPT is found to be evenly distributed between 1 and 3, which means that it is easy for this model to generate strong first-order PTs. Finally, the normalised histogram for the modulus of the complex scalar VEV $w_c \equiv |S|$ is strongly peaked around 100~GeV, while its phase $\alpha \equiv {\rm arg}(S)$ is seen to be distributed almost symmetric about the origin with two peaks at $\pm\pi/4$. 
 
\begin{figure}[] 
\includegraphics[scale=0.52]{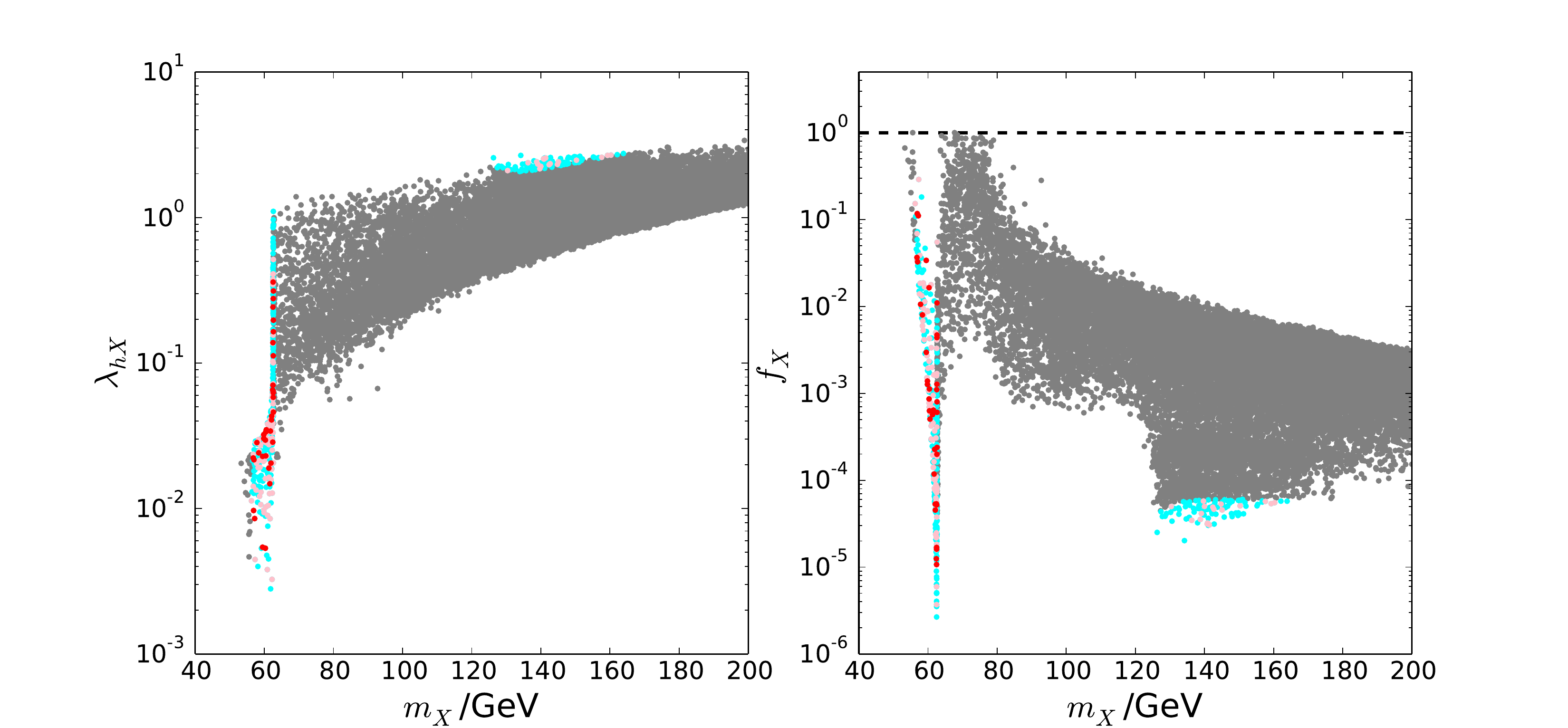}
\caption{The projected parameter space in the $m_X$-$\lambda_{hX}$ plane (left) and in the $m_X$-$f_X$ plane (right). The cyan, pink, red, and blue points are those satisfying the strongly first-order PT conditions and the DM constraints, while the gray points are excluded by the DM direct detection experiment XENON1T. The cyan points are excluded by the conditions $L_w T_c > 3$ and $\alpha<0$, while the pink points by the cutoff scale conditions $\Lambda > 500$~GeV and $w_c^2 /\Lambda^2 < 0.5$. The red points satisfy all the constraints. }\label{mXlhX}
\end{figure}

We now turn to the DM physics in this complex singlet model. We show in Fig.~\ref{mXlhX} the scatter plot of the accepted parameter points in the $m_X$-$\lambda_{hX}$ (left panel) and the $m_X$-$f_X$ (right panel) planes.  
The gray points represent the models ruled out by the DM direct detection experiment XENON1T, while the other color (cyan+pink+red) points denote those consistent with the DM constraints and the baryon-asymmetry washout bound in Eq.~(\ref{xi}). As a result, it is observed that the aforementioned two DM mass regions actually correspond to two different mechanisms to generate the DM relic density. When the DM mass $m_X$ is in the narrow region (I), the DM annihilation during its thermal freeze-out is enhanced greatly by the SM Higgs resonance effect, even though the DM-Higgs coupling $\lambda_{hX}$ is always smaller than 1 and can be as small as $10^{-3}$. In contrast, the models in the DM mass region (II) yield their subdominant DM relic densities with $f_X \sim {\cal O}(10^{-5})$ by taking $\lambda_{hX}$ larger than 1. 

\begin{figure}[] 
\includegraphics[scale=0.58]{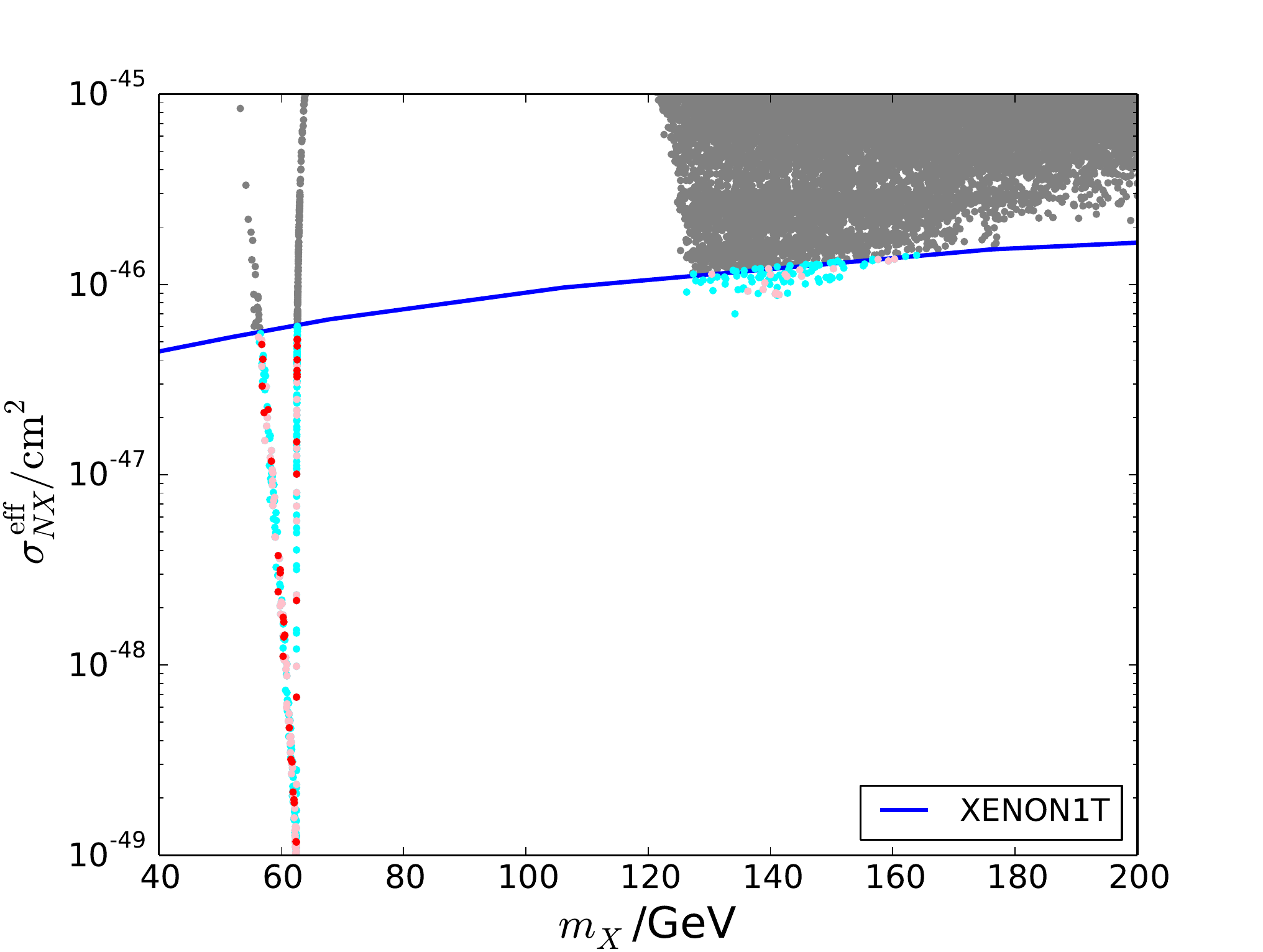}
\caption{The projected parameter space in the $m_X$-$\sigma^{\rm eff}_{NX}$ plane. The color coding of the points is the same as that in Fig.~\ref{mXlhX}. The blue solid curve represents the most recent XENON1T upper limits on the spin-independent DM-nucleon cross sections. }\label{mXsigNX}
\end{figure}

Moreover, we find that the strongest constraint on the DM properties is given by the XENON1T upper limits on the spin-independent DM-nucleon cross sections, which is clearly shown in $m_X$ versus $\sigma^{\rm eff}_{XN}$ plot in Fig.~\ref{mXsigNX}.  The experiments from the DM indirect detections and collider searches do not provide any additional useful constraints to the models.
This feature can be understood as follows. The DM physics in the present model is essentially controlled by two parameters, the DM mass $m_X$ and its Higgs portal coupling $\lambda_{hX}$. For a given DM mass, all of the experimental constraints can only limit $\lambda_{hX}$. Due to the extreme accuracy of DM direct detections, other kinds of experiments cannot provide competitive sensitivity.

Finally, note that none of models left in our random scan can give rise to the observed DM relic abundance. As shown in the $m_X$-$f_X$ plane in Fig.~\ref{mXlhX}, the fraction of DMs in the high-mass region (II) is constrained to be less than $10^{-4}$ by the DM direct detections, while the DMs in the Higgs-resonance region (I) can have the fraction of ${\cal O}(0.1)$ even with relatively small couplings $\lambda_{hX} \lesssim 1$.


\section{Electroweak Baryogenesis}\label{BAU}
In the present model, when the complex scalar $S$ acquires a complex VEV $\langle S\rangle = w_c e^{i\alpha}/\sqrt{2}$, the $CP$ symmetry is broken spontaneously, which, assisted by the dimension-6 effective operator ${\cal O}_6$ in Eq.~(\ref{O6}) can generate a new complex-valued contributions to the top-quark Yukawa coupling
\begin{eqnarray}\label{NYukawa}
\frac{w_c^2 e^{i 2\alpha}}{2\Lambda^2} \bar{Q}_{3L} \tilde{H} t_R + {\rm H.c.} \,.
\end{eqnarray}

The first-order EWPT proceeds via the nucleation of EW-symmetry-breaking bubbles in the high-temperature EW symmetric phase. Due to the spontaneous breaking of the $Z_2$ and $CP$ symmetries in the EW symmetric phase, it is expected that the whole Universe can be divided by many domains characterised by four distinct vacua, $\langle S \rangle = \pm w_c e^{\pm i\alpha}/\sqrt{2}$, each of which occupy the same volume~\cite{McDonald:1993ey,McDonald:1995hp,Comelli:1993ne}. However, it is evident from Eq.~(\ref{NYukawa}) that the CPV phase induced by the vacua $\pm w_c e^{i\alpha}/\sqrt{2}$ is opposite of that in the vacua $\pm w_c e^{-i\alpha}/\sqrt{2}$. Thus, the yielded baryon asymmetries obtained in these two pairs of vacua should be also opposite, and would annihilate with each other when different bubbles collide. As a result, the net baryon asymmetry left after the EW phase transition would vanish. One simple way to avoid such annihilation of baryon asymmetry is to introduce an explicit CPV phase in the scalar potential Eq.~(\ref{V0})~\cite{McDonald:1993ey,McDonald:1995hp,Comelli:1993ne}, which will be discussed in Sec.~\ref{dw}. In this and next sections, we only consider the baryon asymmetry obtained with the bubble nucleation from one specific EW symmetric vacuum with $\langle S \rangle =  w_c e^{i\alpha}/\sqrt{2}$. Note that the effective operator ${\cal O}_6$ respects the $Z_2$ symmetry, so that the CPV effects in Eq.~(\ref{NYukawa}) from $\pm w_c e^{i\alpha}/\sqrt{2}$ are the same. This indicates that the the phase $\alpha$ can be restricted between $-\pi/2$ and $\pi/2$ without any loss of generality.

With the new complex contribution to the top-quark Yukawa coupling in Eq.~(\ref{NYukawa}), the top quark mass inside the bubble wall becomes spatially varying, which is given by
\begin{eqnarray}\label{Mt}
m_t (z) = \frac{y_t}{\sqrt{2} }h(z) \left(1+\frac{ S(z)^2}{y_t \Lambda^2 }\right) \equiv |m_t(z)|e^{i\theta(z)}\,,
\end{eqnarray}
where $S(z)$ and $h(z)$ denote the field profiles of $S$ and the SM Higgs around the bubble wall with $z$ the coordinate transverse to it. Here we assume that the bubble wall has already been large enough so that we can ignore the wall curvature and approximate it as planar. 

Now we follow the procedure given in Ref.~\cite{Espinosa:2011eu} to approximate the bubble wall profile analytically. Firstly, we assume that the field configurations in the vicinity of the wall is given by
\begin{eqnarray}\label{BW}
S(z) &\equiv& \frac{w_c e^{i\alpha}}{2\sqrt{2}} [1+\tanh(z/L_w)]\,,\\
h(z) &\equiv& \frac{v_c}{2} [1-\tanh(z/L_w)]\,,
\end{eqnarray}
where $L_w$ represents the width of the bubble wall. Next we approximate the wall width with the thin-wall approximation. The tunnelling path can be obtained by extremizing the following Euclidean action~\cite{Espinosa:2011eu}
\begin{eqnarray}\label{Euclidean}
S_E = \int^\infty_{-\infty} d\tau \left[\frac{1}{2} (\partial_\tau h)^2 + \frac{1}{2} (\partial_\tau s)^2 + (\partial_\tau a)^2 + V_T(h,s,a) \right]\,,
\end{eqnarray}
with the boundary conditions
\begin{eqnarray}\label{BC}
&&h(-\infty) = v_c\,, \quad h(\infty) = 0\,, \quad h^\prime(\pm\infty) = 0\,,\nonumber\\
&&s(-\infty) = 0\,, \quad s(\infty) =  w_c \cos\alpha\,,\quad s^\prime(\pm\infty) = 0\,,\nonumber\\
&&a(-\infty) = 0\,, \quad a(\infty) =  w_c\sin\alpha\,, \quad a^\prime(\pm\infty) =0\,.
\end{eqnarray}
where $V_T$ is the  finite-temperature effective scalar potential in Eq.~(\ref{V_tot}). We have rewritten $S = (s+ia)/\sqrt{2}$ as its real and imaginary components, and the asymptotic values of $s$ and $a$ approach to their values at the vacua $\langle S\rangle $. It is expected that the final path would pass or be very close to the scalar potential saddle point, whose potential value is given by
\begin{eqnarray}
V_\times = \frac{N_\times}{D_\times}\,,
\end{eqnarray}
where 
\begin{eqnarray}
N_\times &=& v_c^4 w_c^2 \Big(\kappa_{ha}+\kappa_{hs} + (\kappa_{hs}-\kappa_{ha})\cos(2\alpha)\Big)^2 \Big( 128 \lambda_{hs}\lambda_{ha}v_c^4 + 3(\lambda_{hs}+\lambda_{ha})\lambda_{sa}w_c^4 \nonumber\\
&& + (\lambda_{hs}+\lambda_{ha})\lambda_{sa} w_c^4 \big(\cos(8\alpha)-4\cos(4\alpha)\big) \Big)\,,\nonumber\\
D_\times &=& 4096\lambda_{hs}\lambda_{ha}(\kappa_{hs}+\kappa_{ha})v_c^6 + 768(\kappa_{hs}^2 \lambda_{ha} + \kappa_{ha}^2 \lambda_{hs})v_c^4 w_c^2 \nonumber\\
&& +96(\kappa_{hs}+\kappa_{ha})(\lambda_{ha}+\lambda_{hs})\lambda_{sa} v_c^2 w_c^4 + 2\lambda_{sa}(7\kappa_{ha}^2 + 10\kappa_{ha}\kappa_{hs} + 7 \kappa_{hs}^2)w_c^6 \nonumber\\
&&-8\cos(2\alpha)\Big(512(\kappa_{ha}-\kappa_{hs})\lambda_{hs}\lambda_{ha}v_c^6 + 128(\kappa_{ha}^2 \lambda_{hs}-\kappa_{hs}^2 \lambda_{ha})v_c^4 w_c^2 \nonumber\\
&& + 4(\kappa_{ha}-\kappa_{hs})(\lambda_{ha}+\lambda_{hs})\lambda_{sa}v_c^2 w_c^4 + (\kappa_{ha}^2-\kappa_{hs}^2)\lambda_{sa}w_c^6  \Big) \nonumber\\
&& -w_c^2 \cos(4\alpha)\Big(-256(\kappa_{hs}^2\lambda_{ha} + \kappa_{ha}^2 \lambda_{hs})v_c^4 + 128(\kappa_{ha}+\kappa_{hs})(\lambda_{ha}+\lambda_{hs})\lambda_{sa}v_c^2 w_c^2 \nonumber\\
&& +(17\kappa_{ha}^2 + 30 \kappa_{ha}\kappa_{hs}+17\kappa_{hs}^2)\lambda_{sa}w_c^4 \Big)\nonumber\\
&& +\lambda_{sa}w_c^4\Big(12\cos(6\alpha)(\kappa_{ha}-\kappa_{hs})\big( 4(\lambda_{ha}+\lambda_{hs})v_c^2 + (\kappa_{hs}+\kappa_{ha})w_c^2 \big)\nonumber\\
&& + \cos(8\alpha)\big( 32(\kappa_{ha}+\kappa_{hs})(\lambda_{ha}+\lambda_{hs})v_c^2 + 2(\kappa_{ha}^2+6\kappa_{ha}\kappa_{hs} +\kappa_{hs}^2)w_c^2 \big)\nonumber\\
&& + 8\cos(10\alpha)(\kappa_{ha}-\kappa_{hs})\big( 4(\lambda_{ha}+ \lambda_{hs})v_c^2 + (\kappa_{ha}+\kappa_{hs})w_c^2 \big)\nonumber\\
&& - 2\cos(12\alpha)(\kappa_{ha}-\kappa_{hs})^2 w_c^2 \Big)\,.
\end{eqnarray}
Note that, if we replace the coordinate $z$ with the Euclidean time $\tau$, Eq.~(\ref{BW}) satisfies the boundary conditions in Eq.~(\ref{BC}), which means that it is a good estimation of the true solution to the tunnelling path. Therefore, the parametric dependence of $L_w$ of the Euclidean action $S_E$ can be estimated as follows:
\begin{eqnarray}\label{EApp}
S_E = \frac{1}{6 L_w}(v_c^2 + w_c^2) + L_w V_\times\,,
\end{eqnarray}
where the first term is obtained by putting solution of Eq.~(\ref{BW}) into the kinetic terms of three scalars in the action Eq.~(\ref{Euclidean}), while the second term is from the potential term by taking into account that the dominant contribution comes from the potential barrier part within the spatial extension of $L_w$. By extremizing the action in Eq.~(\ref{EApp}), we can obtain the following approximate expression for $L_w$  
\begin{eqnarray}\label{Lw}
L_w = \frac{v_c^2 + w_c^2}{6V_\times}\,.
\end{eqnarray}
We have checked that this bubble wall width formula is consistent with the real scalar one in Ref.~\cite{Espinosa:2011eu}.

It is shown in Ref.~\cite{Cline:2012hg,Fromme:2006wx,Joyce:1994zt,Cline:1997vk,Cline:2000nw} that the spatially-varying top mass in Eq.~(\ref{Mt}) would generate CPV sources on the top and anti-top quarks when they pass through the wall. The produced CP violation on the wall would transport to the region far inside the symmetric phase, where it biases the anomalous EW sphaleron process to produce the baryon asymmetry. In the literature, this picture is realized by solving the transport equations for chemical potentials $\mu_i$ and velocity perturbations $u_i$ of various SM particles $i$~\cite{Joyce:1994zt,Cline:1997vk,Cline:2000nw}. In particular, the most relevant SM particles in our case involve the left-handed top $t_L$, the left-handed bottom $b_L$, the right-handed top $t_R$. In our work, we make use of the transport equations derived in Ref.~\cite{Fromme:2006wx}, which were obtained with the semiclassical baryogenesis framework~\cite{Joyce:1994zt,Cline:1997vk,Cline:2000nw}. We solve these transport equations by the shooting method~\cite{NRecipe}, and obtain the left-handed baryon chemical potential with the following formula:
\begin{eqnarray}
\mu_{B_L} = \frac{1}{2} (1+4K_{1,t_L})\mu_{t_L} + \frac{1}{2}(1+4K_{1,b_L})\mu_{b_L} +2 K_{1,t_R} \mu_{t_R}\,, 
\end{eqnarray}
where the definitions of the coefficients $K_{1,i}\left(m_i(z)/T\right)$ are given in Ref.~\cite{Fromme:2006wx,Cline:2011mm}. After integrating $\mu_{B_L}$ over the symmetric phase with $z>0$, the baryon asymmetry is given by
\begin{eqnarray}\label{etaB}
\eta_B = \frac{n_B}{s} = \frac{405\Gamma_{\rm sph}}{4\pi^2 v_w g_* T} \int^\infty_0 dz \mu_{B_L}(z) e^{-45 \Gamma_{\rm sph}|z|/(4v_w)}\,,
\end{eqnarray} 
where $\Gamma_{\rm sph} \simeq 10^{-6} T$ is the anomalous sphaleron rate in the EW symmetric phase~\cite{DOnofrio:2014rug}, and $g_* = 106.75$ is the effective degrees of relativistic freedom in the plasma. Here we take the bubble wall velocity to be $v_w = 0.1$. It is shown~\cite{Cline:2012hg} that the predicted baryon asymmetry does not depend on the value of $v_w$ in the range $0.01\leq v_w \leq 0.1$, since for a small $v_w$, $\mu_{B_L} \propto v_w$ which is cancelled by the factor $v_w$ in the denominator of Eq.~(\ref{etaB}). 
\begin{figure}[]
\includegraphics[scale = 0.45,angle=-90]{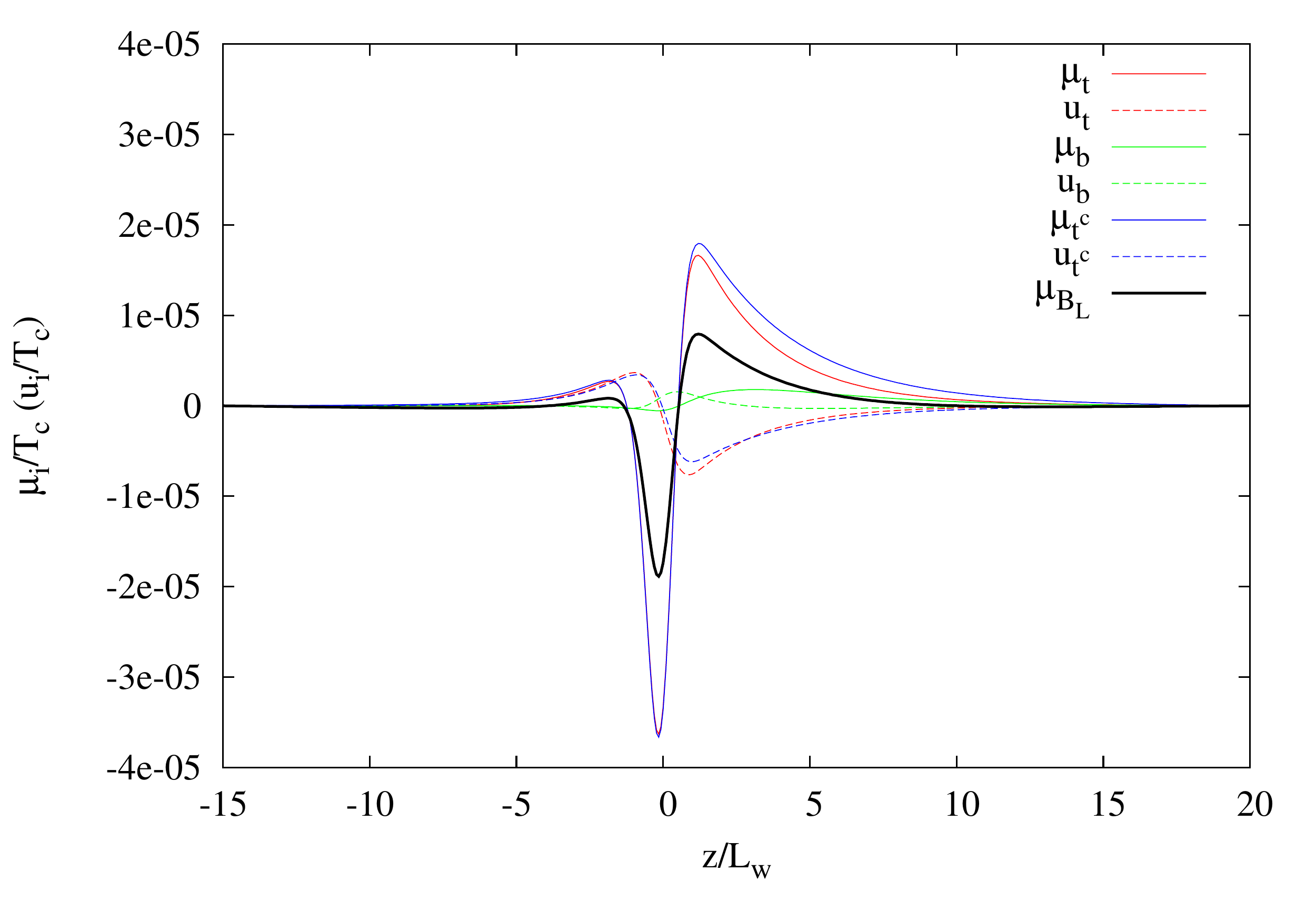}
\caption{An example of solutions to the transport equations and its obtained left-handed baryon chemical potential $\mu_{B_L}$ as functions of the coordinate $z$ transverse to the bubble wall. All of the chemical potentials $\mu_i$ and velocities $u_i$ are noralized with respect to the critical temperature $T_c$, while the coordinate $z$ is normalized with respect to the bubble wall width $L_w$. The corresponding parameters are $v_c = 155.9$~GeV, $w_c = 525.3$~GeV, $\alpha = -1.225$, $T_c =93.85$~GeV, $L_w T_c = 5.111$ and $\Lambda = 1131$~GeV, which can give the observed baryon asymmetry.}\label{muB}
\end{figure}
Fig.~\ref{muB} shows one prototypical solution to the transport equations, as well as its predicted left-handed baryon chemical potential $\mu_{B_L}$, which can give rise to a baryon asymmetry equal to the observed value.

\begin{figure}[]
\includegraphics[scale=0.6]{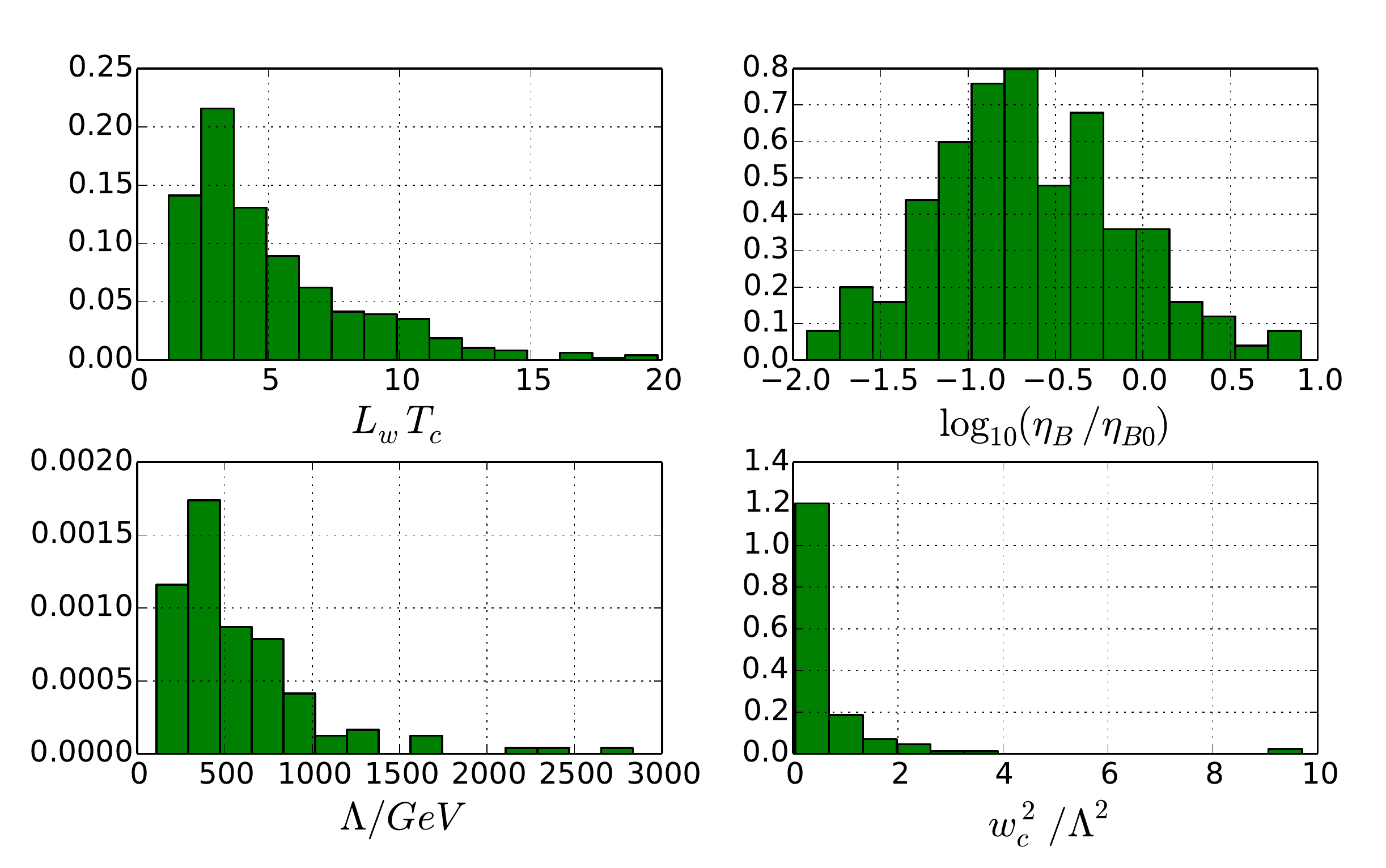}
\caption{Distributions of the bubble wall thickness $L_w$ multiplied by $T_c$, the logarithm of baryon asymmetry $\eta_B$ normalized by its observed value $\eta_{B0}$ with a fixed cutoff scale $\Lambda=1$~TeV, the cutoff $\Lambda$ by rescale the $\eta_B$ to its observed value, and $w^2_c/\Lambda^2$.   }\label{histBAU}
\end{figure}

\begin{figure}[]
\includegraphics[scale=0.46]{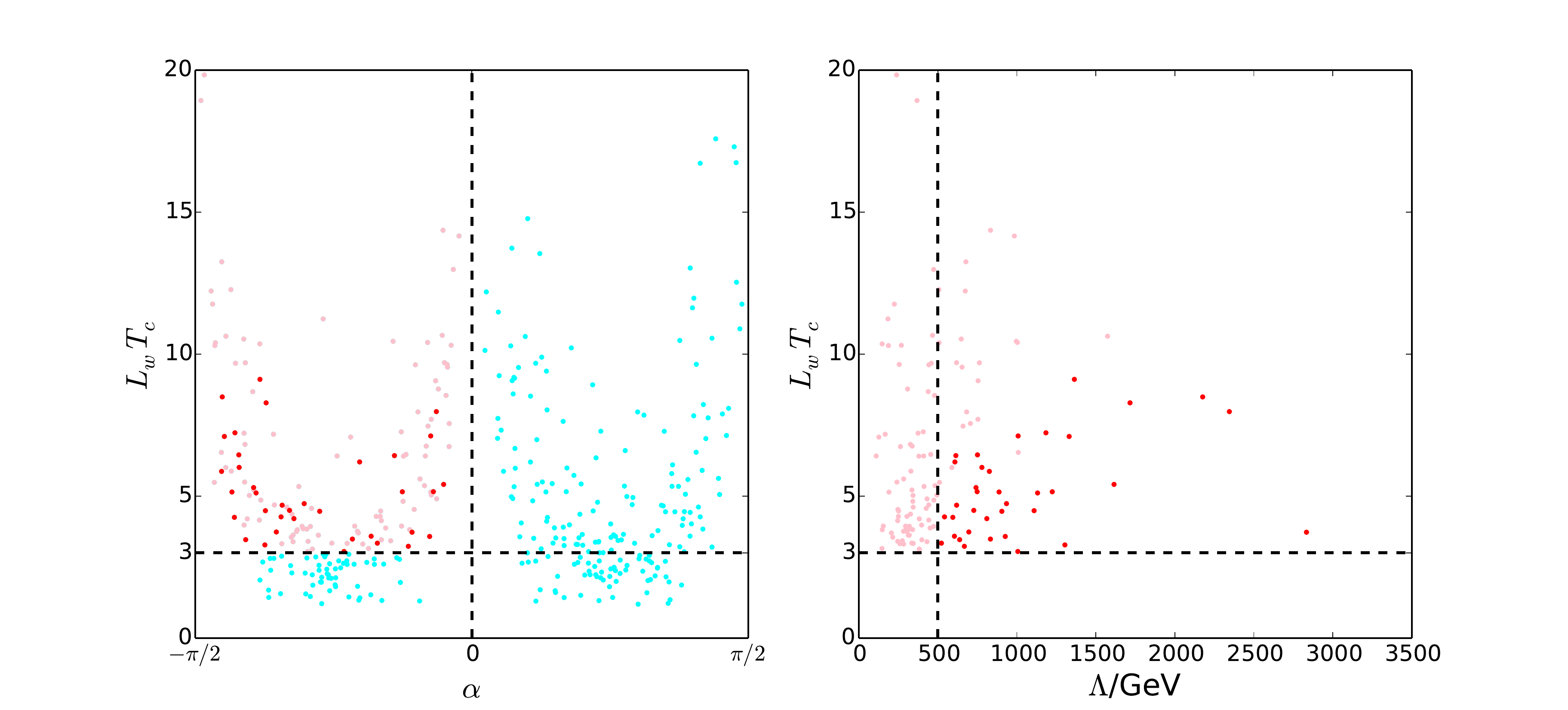}
\caption{Scatter plots in CPV phase $\alpha$ vs. bubble wall width $L_w T_c$ (left) and in $\Lambda$ vs. $L_w T_c$ (right) for the models which satisfy all of the DM and strong first-order PT constraints. The cyan points are excluded by the validity of the semiclassical framework $L_w T_c \geq 3$ and the correct sign of baryon asymmetry $\alpha<0$, while the pink and red points satisfy both conditions. The pink points are further ruled out by the large cutoff condition $\Lambda>500$~GeV and $w_c^2/\Lambda^2 < 0.5$. }\label{alphaLw}
\end{figure}

Of the models passing through all of the DM and strong first-order PT constraints in Sec.~\ref{DMPheno}, they should satisfy two further conditions. Note that the transport equations are derived with the semiclassical framework, so that the consistency requires that $L_w \gg 1/T_c$. In the literature, it is usually assumed to have $L_w T_c \geqslant 3$~\cite{Cline:2012hg}. The distribution of $L_w T_c$ for all the surviving models is shown in the upper left plot of Fig.~\ref{histBAU}, from which we know that nearly half of the models can be allowed by the above bound. Also, it is found that only the negative $\alpha$ can give rise to the correct sign of baryon asymmetry. Therefore, we firstly pick up the models consistent with both conditions, which are shown in the left panel of Fig.~\ref{alphaLw} as the points with red and pink colors.

For the selected models, we calculate the baryon asymmetry produced during the first-order EWPT. We first calculate the baryon asymmetry $\eta_B$ for each model by taking the cutoff scale in ${\cal O}_6$ to be fixed at $\Lambda = 1$~TeV. We display the distribution of the yielded $\eta_B$ in units of the observed value $\eta_{B0} = 8.61\times 10^{-11}$ in the upper right plot of Fig.~\ref{histBAU}, which shows that nearly 10$\%$ of models surpass $\eta_{B0}$. 

We can represent the same information by rescaling the cutoff scale $\Lambda$ so that $\eta_B$ corresponds to its measured value. The distribution of the obtained cutoff scales is shown in the lower right plot of Fig.~\ref{histBAU}. Note that the cutoff scale $\Lambda$ cannot be arbitrarily small for the reliable use of ${\cal O}_6$. Here we restrict $\Lambda>500$~GeV~\cite{Cline:2012hg}, which also singles out about 15$\%$ models. Moreover, we find that large values of the baryon asymmetry are mostly positively correlated to large values of $w_c$, which can be easily understood from Eqs.~(\ref{Mt}) and (\ref{BW}) in that the top quark mass contribution from ${\cal O}_6$ is proportional to $w_c^2$. In order that the dimension-6 operator does not change the top quark mass too much compared with the SM Yukawa couplings~\cite{Cline:2012hg}, we further make the additional constraint $w_c^2/\Lambda^2 < 0.5$. The distribution of $w_c^2/\Lambda^2$ in the lower right plot of Fig.~\ref{histBAU} demonstrates that most models do satisfy this limit. The right panel of Fig.~\ref{alphaLw} shows the constraining power of these two conditions in the $\Lambda$-$L_w T_c$ plane, which indicates that it is relatively easy for the present model to generate the observed baryon asymmetry. We also illustrate the impact of the EW baryogenesis constraints on the DM parameter space in Figs.~\ref{mXlhX} and \ref{mXsigNX}. It is evident that the models which are capable of explaining the cosmological matter-antimatter asymmetry are only located in the Higgs resonance region (I), while the DM high-mass region (II) is completely ruled out.

\section{Models with the Correct Dark Matter Relic Density}\label{GoodModel}
In previous sections, after performing the large-scale scan of parameter space, only when the DM mass is nearly half of the SM Higgs mass can we find the models to accommodate baryon asymmetry. Unfortunately, all of allowed models cannot give rise to the observed DM relic density. 
Therefore, the problem we are concerned with in this section is if it is possible to find models which can explain the observed DM relic density and baryon asymmetry simultaneously while they are consistent with all experimental constraints.
In order to achieve this, we make a dedicated parameter scan by fixing the DM mass in the Higgs-resonance region $m_X = 55\sim 65$~GeV and allowing the DM relic abundance within the $1\sigma$ range of the Planck measured value $\Omega_{\rm DM} h^2 = 0.1186\pm 0.0020$~\cite{Ade:2015xua}. We also restrict the DM to be the pseudoscalar $a$ without loss of generality since $s$ and $a$ are equivalent in the scalar potential in Eq.~(\ref{V0}). Furthermore, the CPV phase is required to be in the range $-\pi/2 \leq \alpha \leq 0$ in order to achieve the correct sign of baryon asymmetry, and the bubble wall width satisfies $L_w T_c \geq 3$ for the validity of semiclassical treatment of the transport equations. 

\begin{figure}[]
\includegraphics[scale=0.55]{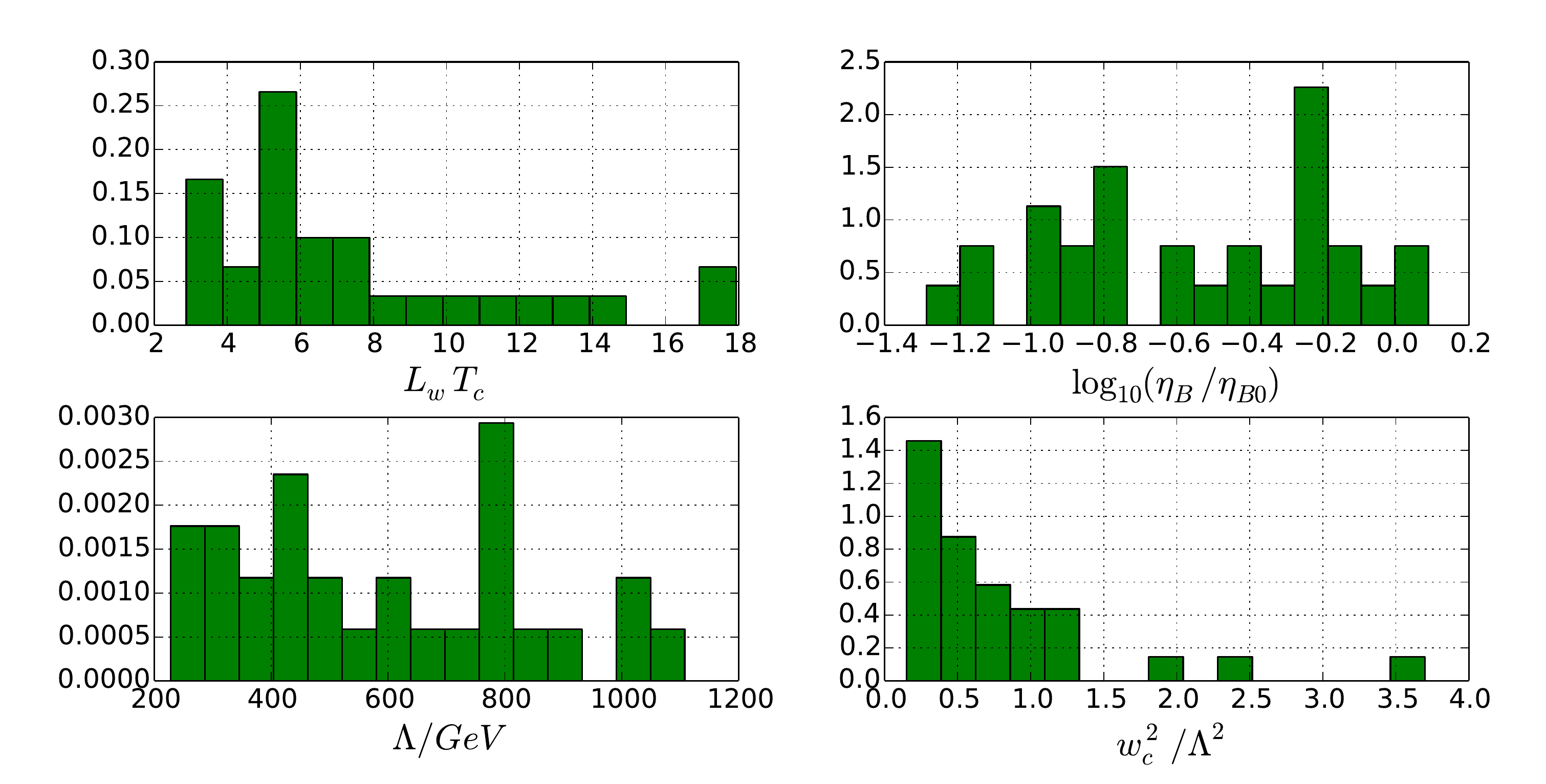}
\caption{The same distributions as in Fig.~\ref{histBAU} but for the second scan of models which tries to explain the DM relic abundance and the baryon asymmetry in the Universe simultaneously. }\label{histS}
\end{figure}

\begin{figure}[th]
\includegraphics[scale=0.5]{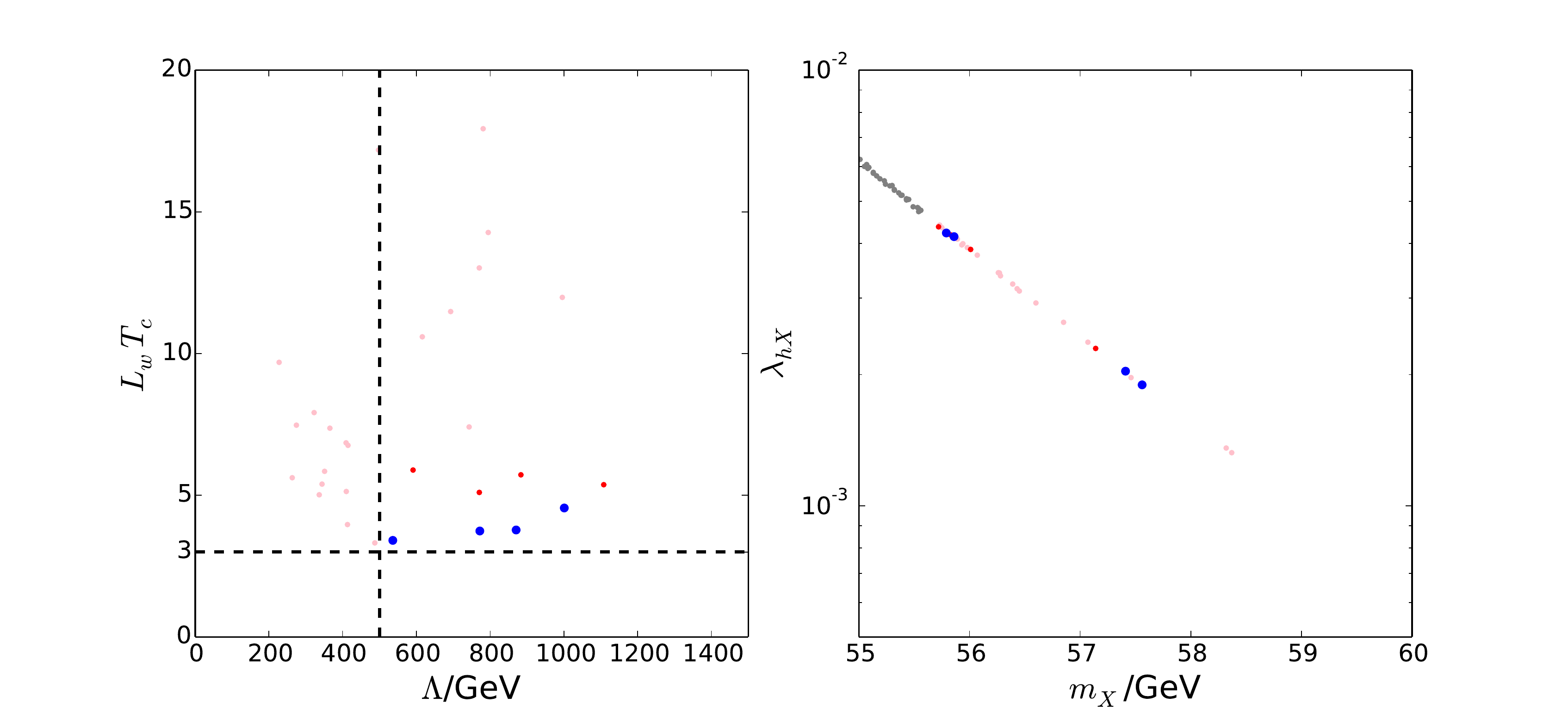}
\caption{Scatter plots of $\Lambda$ vs. $L_w T_c$ (left) and $m_X$ vs. $\lambda_{hX}$ (right) according to the data obtained by the second scan. The pink, red and blue points represent models which can explain the DM relic abundance and the baryon asymmetry at the same time. The red and blue points are allowed by $\Lambda > 500$~GeV and $w_c^2/\Lambda^2<0.5$ for the validity of the effective operator ${\cal O}_6$, while the pink points are excluded. The bigger blue points represent the benchmark models which can even satisfy $w_c^2/\Lambda^2 < 0.2$. The gray points on the right plot shows those which are ruled out by the current DM direct detection upper bounds.  
 }\label{LambLwS}
\end{figure}

As a result, with the scanning of about $2\times 10^7$ models, we can find 30 models in total to satisfy all of the above requirements. For the remaining models, we then calculate the baryon asymmetry for each of them. Following Sec.~\ref{BAU}, the results can be represented in terms of either the predicted baryon asymmetry $\eta_B$ by fixing $\Lambda=1~$TeV or the cutoff scale $\Lambda$ by fixing the asymmetry to be the observed one. The final distributions of various physical quantities are shown in Figs.~\ref{histS}.  
If we further impose the conditions $\Lambda>500$~GeV and $w_c^2/\Lambda^2<0.5$ to guarantee the appropriate use of the effective operator ${\cal O}_6$, we finally select 8 models which can meet these two extra limits. The distribution of these points are shown as the red and blue points in the left panel of Fig.~\ref{LambLwS}, from which it is seen that these conditions favor the models with relatively small bubble wall with $L_w T_c \lesssim 8$. We also plot as bigger blue dots the models which can be in accord with more stringent constraints $w_c^2/\Lambda^2 < 0.2$, which further reduce the wall width to $L_w T_c \lesssim 5$. 
We can also show the baryon asymmetry constraints on the DM properties in the selected models by making the plot in the plane of the DM mass $m_X$ vs. its Higgs portal coupling $\lambda_{hX}$, which is given by the right panel of Fig.~\ref{LambLwS}. As a result, the DM mass is predicted in the small range of $55.5~{\rm GeV} \lesssim m_X \lesssim 58$~GeV, and the coupling to be $\lambda_{hX}\sim {\cal O}(10^{-3})$. It is interesting to note that the allowed DM mass is a little smaller than the half of the SM Higgs mass $m_h = 62.5$~GeV. In this situation, the DM thermal kinetic energies at the freeze-out time increase the center-of-mass energy of two-DM system so that the total energy approaches to the SM Higgs pole mass more closely, which makes the Higgs resonance enhancement more pronounced. It turns out that the DM-Higgs coupling $\lambda_{hX}$ can be reduced greatly, which is further helpful for the model to escape the severe DM direct detection bound. 

\section{Domain Walls and Explicit CP Violation}\label{dw}
In the previous sections, we simply assumed that at the time just before the EWPT, the Universe should be filled with only one EW symmetric vacuum with $(0,w_c e^{i\alpha}/\sqrt{2})$ where the two entries represent the VEVs of the SM Higgs and the complex scalar, respectively. However, the Universe should experience a two-step PT in the present model. In the first step, the EW symmetry is kept while the VEV of $S$ breaks $Z_2$ and $CP$ symmetries in the scalar potential. Thus, there is expected to be four distinct vacua, parametrized by $\langle S \rangle = \pm w_d e^{\pm i\alpha}$. Here we take $w_d$ to be positive, which should be distinguished from its critical temperature value $w_c$. We also define the phase in the range $0\leq \alpha \leq \pi/2$, which is different from that in the previous sections. These four vacua are expected to occupy the same spatial volume in the whole Universe, and they are separated by different kinds of domain walls. When the temperature drops down to the EW critical temperature $T_c$, the EWPT occurs when the bubbles of the unique EW breaking phase with $(v_c,0)$ begin to nucleate inside every vacuum patch. On the other hand, the transitions from the vacua $(0, \pm w_c e^{i\alpha}/\sqrt{2})$ should generate the negative value of the baryon asymmetry, which is in contrast to the vacua $(0, \pm w_c e^{-i\alpha}/\sqrt{2})$ leading to the positive baryon number. Eventually, when all of the EW breaking bubbles collide, the produced baryon asymmetries in different patches would be neutralized with each other due to their opposite signs. Therefore, it is generically regarded that the models with exact $CP$ symmetry cannot generate the net baryon asymmetries~\cite{McDonald:1993ey,McDonald:1995hp,Comelli:1993ne}. 

In the literature, one easy way to avoid such an exact baryon-number cancellation is to introduce a explicit CPV phase in the scalar potential $V_0$ to dynamically remove the vacua $(0,\pm w_c e^{i\alpha}/\sqrt{2})$ with the wrong sign of baryon asymmetry~\cite{McDonald:1993ey, McDonald:1995hp, Comelli:1993ne}. We follow this line of thinking in the present section and focus on the case in which the explicit $CP$ violation takes place through the quartic term $S^4$. In particular, we try to estimate the required size of the corresponding CPV phase. Note that $S^4$ appears only through the term $\lambda_2 (S^2 + S^{*2})^2/4$ in Eq.~(\ref{V0}). Thus, we would like to rewrite this term in the following way
\begin{eqnarray}
V_4 = \frac{\lambda_2 e^{i\delta}}{4} S^4 + \frac{\lambda_2 e^{-i\delta}}{4} S^{*4} + \frac{\lambda_2}{2} |S|^4\,,  
\end{eqnarray}
in which we have introduced a small phase $\delta$ while still kept $\lambda_2$ as a real parameter. With this explicit CP violation, the vacua $(0,\pm w_d e^{i\alpha}/\sqrt{2})$ at $T>T_c$ would have the potential density
\begin{eqnarray}
V_T^{+} = \frac{1}{8}\lambda_2 w_d^4 \cos(\delta+4\alpha) + V_T^{\rm CP}\,,
\end{eqnarray}
while the potential density for $(0,\pm w_d e^{-i\alpha}/\sqrt{2})$ is
\begin{eqnarray}
V_T^- = \frac{1}{8} \lambda_2 w_d^4 \cos(\delta-4\alpha) + V_T^{\rm CP}\,,
\end{eqnarray}
where $V_T^{\rm CP}$ denotes other terms which is invariant under the $CP$ transformation. Therefore, the potential difference between two pairs of vacua is given by
\begin{eqnarray}
\Delta V_T = -\frac{1}{4} \lambda_2 w_d^4 \sin(4\alpha) \sin\delta\,.
\end{eqnarray}
If $\Delta V_T > 0$, then it indicates that the vacua $(0, \pm w_d e^{-i\alpha}/\sqrt{2})$ with the right-sign baryon asymmetry is more energetically favored against the wrong-sign vacua $(0, \pm w_d e^{i\alpha}/\sqrt{2})$.

It is shown in Ref.~\cite{Lew:1993yt,McDonald:1995hp} that the disappearance of the wrong-sign vacua can proceed via the movement of the domain walls interpolating between the wrong-sign and right-sign vacua. In this process, the volume originally occupied by the wrong-sign vacua is transferred into the right-sign counterparts. Note that a domain wall begin to move when the energy scale of the potential difference between the adjunct vacua approaches that of its surface energy density $\eta_{\rm DW}$, which is usually of order of $\eta_{\rm DW} \sim w_d^3$. Thus, we can estimate the time for the bubble wall movement as follows:
\begin{eqnarray}
t_{\rm DW} \approx \frac{\eta_{\rm DW}}{|\Delta V_T|} \sim \frac{1}{|\lambda_2\sin(4\alpha)\sin\delta|w_d}\,.
\end{eqnarray}  
The consistency of our picture for EW baryogenesis requires to eliminate the wrong-sign domains at least before the EWPT, which takes place at the time $t_{\rm EW} \sim M_{\rm Pl}/T^2_c$ with $M_{\rm Pl}$ the Planck mass and $T_c$ the critical temperature. Therefore, we should have $t_{\rm DW} < t_{\rm EW}$, which can be translated into the following constraint
\begin{eqnarray}
|\sin\delta| > \frac{T_c^2}{|\lambda_2 \sin(4\alpha)|w_d M_{\rm Pl}} \sim \frac{T_c^2}{|\lambda_2 \sin(4\alpha)|w_c M_{\rm Pl}}\,,
\end{eqnarray}  
where we have approximate $w_d$ with its critical temperature value $w_c$ in the second relation. If we take the PT parameters as their typical values $T_c \sim 100~$GeV, $w_c \sim 100~$GeV, $|\sin(4\alpha)| \sim 0.1$ and $|\lambda_2| \sim {\cal O}(0.1)$, the CPV phase are only needed to be larger than ${\cal O}(10^{-15})$. In other words, as long as the CPV phase is chosen to satisfy this bound, the wrong-sign vacuum domains would shrink rapidly and disappear totally when the associated domain walls collide and annihilate. Obviously, such a small phase cannot provide any visible CPV effects under the current experimental status. The above estimation of the CPV phase in the scalar potential agrees with that in Ref.~\cite{McDonald:1995hp}.

Until now, we have focused on the elimination of the domain walls involving the wrong-sign vacua $\pm w_d e^{i\alpha}/\sqrt{2}$. However, due to the $Z_2$ symmetry breaking, there is still another kind of domain walls which divide the two right-sign vacua $(0,\pm w_d e^{-i\alpha}/\sqrt{2})$. However, it is well known that we do not need to worry about them since they would decay after the $Z_2$ symmetry is restored almost at the EW critical temperature $T_c \sim 100$~GeV, which is well before they could dominate the energy density of the Universe at $T \sim 10^{-7}$~GeV~\cite{Espinosa:2011eu,Cline:2012hg}.  

\section{Conclusions}\label{conclusion}
We have investigated a new connection between the DM physics and the EW baryogenesis in a simple extension of the SM by introducing an additional complex EW singlet scalar $S$ and imposing the $CP$ and $Z_2$ symmetries. On one hand, at the temperature just above the EWPT, $S$ acquires a complex-valued VEV, which generate a tree-level barrier between the EW symmetric and broken phases at the EW critical temperature $T_c$. The EWPT can be of strongly first order, and, assisted by the effective operator ${\cal O}_6$, the $CP$ symmetry is spontaneously broken at finite temperatures, both of which are of great importance to the successful EW baryogenesis. On the other hand, after the EWPT, the $Z_2$ and $CP$ symmetries are restored, so that a DM candidate arises as the lighter component of $S$ which is stabilized by the $Z_2$ symmetry, and the severe constraints on $CP$ violations from low-energy EDM measurements can be evaded. As a result, it has been shown that we can simultaneously generate the observed DM relic density and the baryon asymmetry in the Universe only when the DM mass is in the SM Higgs resonance region and the Higgs portal coupling is of ${\cal O}(10^{-3})$. Furthermore, as for the vacuum domains which produce the excesses of antibaryon number and cancel the baryon asymmetries in the right-sign vacua, we have shown that it is sufficient to introduce a tiny explicit CPV phase of ${\cal O}(10^{-15})$ in the scalar potential so that such wrong-sign domains could disappear before the EWPT. Here we would like to emphasize different roles played by the spontaneous and explicit CPV phases in this scenario. The former is the true source of the $CP$ violation necessary to generate the baryon asymmetry, while the later just lifts the degeneracy in potential between the right-sign and wrong-sign vacua to achieve the net baryon asymmetry. 

Note that our model can easily generate a strong first-order EWPT. It has been argued in Ref.~\cite{Vaskonen:2016yiu,No:2011fi,Caprini:2015zlo} that such a strong PT can also produce a strong gravitational wave signal, which could be detected by the near-future gravitational wave experiments such as LISA~\cite{LISA} or BBO~\cite{BBO} interferometers. It is intriguing that our model can be further tested by the gravitational wave observations, which is, however, beyond the scope of the current work.

\appendix

\section*{Acknowledgments}
The authors thank Jose Wudka for his interest at the beginning of this project. This work is supported by the National Science Centre (Poland), research projects no~2014/15/B/ST2/00108 and no~2017/25/B/ST2/00191.



\begin{thebibliography}{0}
\bibitem{pdg} 
C.~Patrignani {\it et al.} (Particle Data Group), 
Chin.\ Phys.\ C,\ {\bf 40}, 100001 (2016) and 2017 update
doi:10.1088/1674-1137/40/10/100001

\bibitem{Bergstrom:2012fi} 
  L.~Bergstrom,
  Annalen Phys.\  {\bf 524}, 479 (2012)
  doi:10.1002/andp.201200116
  [arXiv:1205.4882 [astro-ph.HE]].
  
\bibitem{Ade:2015xua} 
  P.~A.~R.~Ade {\it et al.} [Planck Collaboration],
  Astron.\ Astrophys.\  {\bf 594}, A13 (2016)
  doi:10.1051/0004-6361/201525830
  [arXiv:1502.01589 [astro-ph.CO]].
  
\bibitem{Rubin:1970zza} 
  V.~C.~Rubin and W.~K.~Ford, Jr.,
  Astrophys.\ J.\  {\bf 159}, 379 (1970).
  doi:10.1086/150317
  
\bibitem{Jee:2007nx} 
  M.~J.~Jee {\it et al.},
  Astrophys.\ J.\  {\bf 661}, 728 (2007)
  doi:10.1086/517498
  [arXiv:0705.2171 [astro-ph]].
  

\bibitem{Sakharov:1967dj} 
  A.~D.~Sakharov,
  Pisma Zh.\ Eksp.\ Teor.\ Fiz.\  {\bf 5}, 32 (1967)
  [JETP Lett.\  {\bf 5}, 24 (1967)]
  [Sov.\ Phys.\ Usp.\  {\bf 34}, no. 5, 392 (1991)]
  [Usp.\ Fiz.\ Nauk {\bf 161}, no. 5, 61 (1991)].
  doi:10.1070/PU1991v034n05ABEH002497

\bibitem{Kuzmin:1985mm} 
  V.~A.~Kuzmin, V.~A.~Rubakov and M.~E.~Shaposhnikov,
  Phys.\ Lett.\  {\bf 155B}, 36 (1985).
  doi:10.1016/0370-2693(85)91028-7
  
\bibitem{Cohen:1990py} 
  A.~G.~Cohen, D.~B.~Kaplan and A.~E.~Nelson,
  Phys.\ Lett.\ B {\bf 245}, 561 (1990).
  doi:10.1016/0370-2693(90)90690-8

\bibitem{Cohen:1993nk} 
  A.~G.~Cohen, D.~B.~Kaplan and A.~E.~Nelson,
  Ann.\ Rev.\ Nucl.\ Part.\ Sci.\  {\bf 43}, 27 (1993)
  doi:10.1146/annurev.ns.43.120193.000331
  [hep-ph/9302210].

\bibitem{Rubakov:1996vz} 
  V.~A.~Rubakov and M.~E.~Shaposhnikov,
  Usp.\ Fiz.\ Nauk {\bf 166}, 493 (1996)
  [Phys.\ Usp.\  {\bf 39}, 461 (1996)]
  doi:10.1070/PU1996v039n05ABEH000145
  [hep-ph/9603208].

\bibitem{Cline:2000fh} 
  J.~M.~Cline,
  Pramana {\bf 55}, 33 (2000)
  doi:10.1007/s12043-000-0081-6
  [hep-ph/0003029].

\bibitem{Morrissey:2012db} 
  D.~E.~Morrissey and M.~J.~Ramsey-Musolf,
  New J.\ Phys.\  {\bf 14}, 125003 (2012)
  doi:10.1088/1367-2630/14/12/125003
  [arXiv:1206.2942 [hep-ph]].
  
\bibitem{Konstandin:2013caa} 
  T.~Konstandin,
  Phys.\ Usp.\  {\bf 56}, 747 (2013)
  [Usp.\ Fiz.\ Nauk {\bf 183}, 785 (2013)]
  doi:10.3367/UFNe.0183.201308a.0785
  [arXiv:1302.6713 [hep-ph]].
  
\bibitem{Klinkhamer:1984di} 
  F.~R.~Klinkhamer and N.~S.~Manton,
  Phys.\ Rev.\ D {\bf 30}, 2212 (1984).
  doi:10.1103/PhysRevD.30.2212
  
\bibitem{Gavela:1994dt} 
  M.~B.~Gavela, P.~Hernandez, J.~Orloff, O.~Pene and C.~Quimbay,
  Nucl.\ Phys.\ B {\bf 430}, 382 (1994)
  doi:10.1016/0550-3213(94)00410-2
  [hep-ph/9406289].
  
\bibitem{Huet:1994jb} 
  P.~Huet and E.~Sather,
  Phys.\ Rev.\ D {\bf 51}, 379 (1995)
  doi:10.1103/PhysRevD.51.379
  [hep-ph/9404302].
  
\bibitem{Kajantie:1996mn} 
  K.~Kajantie, M.~Laine, K.~Rummukainen and M.~E.~Shaposhnikov,
  Phys.\ Rev.\ Lett.\  {\bf 77}, 2887 (1996)
  doi:10.1103/PhysRevLett.77.2887
  [hep-ph/9605288].
  
\bibitem{Csikor:1998eu} 
  F.~Csikor, Z.~Fodor and J.~Heitger,
  Phys.\ Rev.\ Lett.\  {\bf 82}, 21 (1999)
  doi:10.1103/PhysRevLett.82.21
  [hep-ph/9809291].

\bibitem{Aoki:1999fi} 
  Y.~Aoki, F.~Csikor, Z.~Fodor and A.~Ukawa,
  Phys.\ Rev.\ D {\bf 60}, 013001 (1999)
  doi:10.1103/PhysRevD.60.013001
  [hep-lat/9901021].

\bibitem{Shaposhnikov:1987tw} 
  M.~E.~Shaposhnikov,
  Nucl.\ Phys.\ B {\bf 287}, 757 (1987).
  doi:10.1016/0550-3213(87)90127-1

\bibitem{Farrar:1993hn} 
  G.~R.~Farrar and M.~E.~Shaposhnikov,
  Phys.\ Rev.\ D {\bf 50}, 774 (1994)
  doi:10.1103/PhysRevD.50.774
  [hep-ph/9305275].
  
\bibitem{Gavela:1993ts} 
  M.~B.~Gavela, P.~Hernandez, J.~Orloff and O.~Pene,
  Mod.\ Phys.\ Lett.\ A {\bf 9}, 795 (1994)
  doi:10.1142/S0217732394000629
  [hep-ph/9312215].

\bibitem{Konstandin:2003dx} 
  T.~Konstandin, T.~Prokopec and M.~G.~Schmidt,
  Nucl.\ Phys.\ B {\bf 679}, 246 (2004)
  doi:10.1016/j.nuclphysb.2003.11.037
  [hep-ph/0309291].
 

\bibitem{McDonald:1993ey} 
  J.~McDonald,
  Phys.\ Lett.\ B {\bf 323}, 339 (1994).
  doi:10.1016/0370-2693(94)91229-7

\bibitem{McDonald:1995hp} 
  J.~McDonald,
  Phys.\ Lett.\ B {\bf 357}, 19 (1995).
  doi:10.1016/0370-2693(95)00716-X
  
\bibitem{Branco:1998yk} 
  G.~C.~Branco, D.~Delepine, D.~Emmanuel-Costa and F.~R.~Gonzalez,
  Phys.\ Lett.\ B {\bf 442}, 229 (1998)
  doi:10.1016/S0370-2693(98)01253-2
  [hep-ph/9805302].

\bibitem{Barger:2008jx} 
  V.~Barger, P.~Langacker, M.~McCaskey, M.~Ramsey-Musolf and G.~Shaughnessy,
  Phys.\ Rev.\ D {\bf 79}, 015018 (2009)
  doi:10.1103/PhysRevD.79.015018
  [arXiv:0811.0393 [hep-ph]].
  
\bibitem{Profumo:2007wc} 
  S.~Profumo, M.~J.~Ramsey-Musolf and G.~Shaughnessy,
  JHEP {\bf 0708}, 010 (2007)
  doi:10.1088/1126-6708/2007/08/010
  [arXiv:0705.2425 [hep-ph]].
  
\bibitem{Gonderinger:2012rd} 
  M.~Gonderinger, H.~Lim and M.~J.~Ramsey-Musolf,
  Phys.\ Rev.\ D {\bf 86}, 043511 (2012)
  doi:10.1103/PhysRevD.86.043511
  [arXiv:1202.1316 [hep-ph]].
  
\bibitem{Costa:2014qga} 
  R.~Costa, A.~P.~Morais, M.~O.~P.~Sampaio and R.~Santos,
  Phys.\ Rev.\ D {\bf 92}, 025024 (2015)
  doi:10.1103/PhysRevD.92.025024
  [arXiv:1411.4048 [hep-ph]].
  
\bibitem{Coimbra:2013qq} 
  R.~Coimbra, M.~O.~P.~Sampaio and R.~Santos,
  Eur.\ Phys.\ J.\ C {\bf 73}, 2428 (2013)
  doi:10.1140/epjc/s10052-013-2428-4
  [arXiv:1301.2599 [hep-ph]].
  
\bibitem{Jiang:2015cwa} 
  M.~Jiang, L.~Bian, W.~Huang and J.~Shu,
  Phys.\ Rev.\ D {\bf 93}, no. 6, 065032 (2016)
  doi:10.1103/PhysRevD.93.065032
  [arXiv:1502.07574 [hep-ph]].
  
\bibitem{Chao:2017oux} 
  W.~Chao,
  arXiv:1706.01041 [hep-ph].
  
\bibitem{Chiang:2017nmu} 
  C.~W.~Chiang, M.~J.~Ramsey-Musolf and E.~Senaha,
  Phys.\ Rev.\ D {\bf 97}, no. 1, 015005 (2018)
  doi:10.1103/PhysRevD.97.015005
  [arXiv:1707.09960 [hep-ph]].

\bibitem{Moore:1998swa} 
  G.~D.~Moore,
  Phys.\ Rev.\ D {\bf 59}, 014503 (1999)
  doi:10.1103/PhysRevD.59.014503
  [hep-ph/9805264].
  
\bibitem{Patel:2011th} 
  H.~H.~Patel and M.~J.~Ramsey-Musolf,
  JHEP {\bf 1107}, 029 (2011)
  doi:10.1007/JHEP07(2011)029
  [arXiv:1101.4665 [hep-ph]].
  
\bibitem{Baron:2013eja} 
  J.~Baron {\it et al.} [ACME Collaboration],
  Science {\bf 343}, 269 (2014)
  doi:10.1126/science.1248213
  [arXiv:1310.7534 [physics.atom-ph]].
  

\bibitem{Espinosa:1993bs} 
  J.~R.~Espinosa and M.~Quiros,
  Phys.\ Lett.\ B {\bf 305}, 98 (1993)
  doi:10.1016/0370-2693(93)91111-Y
  [hep-ph/9301285].

\bibitem{Choi:1993cv} 
  J.~Choi and R.~R.~Volkas,
  Phys.\ Lett.\ B {\bf 317}, 385 (1993)
  doi:10.1016/0370-2693(93)91013-D
  [hep-ph/9308234].
 
\bibitem{Ham:2004cf} 
  S.~W.~Ham, Y.~S.~Jeong and S.~K.~Oh,
  J.\ Phys.\ G {\bf 31}, no. 8, 857 (2005)
  doi:10.1088/0954-3899/31/8/017
  [hep-ph/0411352].
  
\bibitem{Espinosa:2007qk} 
  J.~R.~Espinosa and M.~Quiros,
  Phys.\ Rev.\ D {\bf 76}, 076004 (2007)
  doi:10.1103/PhysRevD.76.076004
  [hep-ph/0701145].

\bibitem{Ahriche:2007jp} 
  A.~Ahriche,
  Phys.\ Rev.\ D {\bf 75}, 083522 (2007)
  doi:10.1103/PhysRevD.75.083522
  [hep-ph/0701192].

\bibitem{Espinosa:2011ax} 
  J.~R.~Espinosa, T.~Konstandin and F.~Riva,
  Nucl.\ Phys.\ B {\bf 854}, 592 (2012)
  doi:10.1016/j.nuclphysb.2011.09.010
  [arXiv:1107.5441 [hep-ph]].
  
\bibitem{Ahriche:2012ei} 
  A.~Ahriche and S.~Nasri,
  Phys.\ Rev.\ D {\bf 85}, 093007 (2012)
  doi:10.1103/PhysRevD.85.093007
  [arXiv:1201.4614 [hep-ph]].

\bibitem{Profumo:2014opa} 
  S.~Profumo, M.~J.~Ramsey-Musolf, C.~L.~Wainwright and P.~Winslow,
  Phys.\ Rev.\ D {\bf 91}, no. 3, 035018 (2015)
  doi:10.1103/PhysRevD.91.035018
  [arXiv:1407.5342 [hep-ph]].
 
\bibitem{Alanne:2014bra} 
  T.~Alanne, K.~Tuominen and V.~Vaskonen,
  Nucl.\ Phys.\ B {\bf 889}, 692 (2014)
  doi:10.1016/j.nuclphysb.2014.11.001
  [arXiv:1407.0688 [hep-ph]].
  
\bibitem{Alanne:2016wtx} 
  T.~Alanne, K.~Kainulainen, K.~Tuominen and V.~Vaskonen,
  JCAP {\bf 1608}, no. 08, 057 (2016)
  doi:10.1088/1475-7516/2016/08/057
  [arXiv:1607.03303 [hep-ph]].

\bibitem{Tenkanen:2016idg} 
  T.~Tenkanen, K.~Tuominen and V.~Vaskonen,
  JCAP {\bf 1609}, no. 09, 037 (2016)
  doi:10.1088/1475-7516/2016/09/037
  [arXiv:1606.06063 [hep-ph]].
  
\bibitem{Vaskonen:2016yiu} 
  V.~Vaskonen,
  Phys.\ Rev.\ D {\bf 95}, no. 12, 123515 (2017)
  doi:10.1103/PhysRevD.95.123515
  [arXiv:1611.02073 [hep-ph]].

\bibitem{Espinosa:2011eu} 
  J.~R.~Espinosa, B.~Gripaios, T.~Konstandin and F.~Riva,
  JCAP {\bf 1201}, 012 (2012)
  doi:10.1088/1475-7516/2012/01/012
  [arXiv:1110.2876 [hep-ph]].

\bibitem{Cline:2012hg} 
  J.~M.~Cline and K.~Kainulainen,
  JCAP {\bf 1301}, 012 (2013)
  doi:10.1088/1475-7516/2013/01/012
  [arXiv:1210.4196 [hep-ph]].
  
\bibitem{Comelli:1993ne} 
  D.~Comelli, M.~Pietroni and A.~Riotto,
  Nucl.\ Phys.\ B {\bf 412}, 441 (1994)
  doi:10.1016/0550-3213(94)90511-8
  [hep-ph/9304267].
  

\bibitem{Haber:2012np} 
  H.~E.~Haber and Z.~Surujon,
  Phys.\ Rev.\ D {\bf 86}, 075007 (2012)
  doi:10.1103/PhysRevD.86.075007
  [arXiv:1201.1730 [hep-ph]].

\bibitem{Nebot:2007bc} 
  M.~Nebot, J.~F.~Oliver, D.~Palao and A.~Santamaria,
  Phys.\ Rev.\ D {\bf 77}, 093013 (2008)
  doi:10.1103/PhysRevD.77.093013
  [arXiv:0711.0483 [hep-ph]].

\bibitem{Belanger:2001fz} 
  G.~Belanger, F.~Boudjema, A.~Pukhov and A.~Semenov,
  Comput.\ Phys.\ Commun.\  {\bf 149}, 103 (2002)
  doi:10.1016/S0010-4655(02)00596-9
  [hep-ph/0112278].
  
\bibitem{Belanger:2013oya} 
  G.~Belanger, F.~Boudjema, A.~Pukhov and A.~Semenov,
  Comput.\ Phys.\ Commun.\  {\bf 185}, 960 (2014)
  doi:10.1016/j.cpc.2013.10.016
  [arXiv:1305.0237 [hep-ph]].




  
\bibitem{Cline:2013gha} 
  J.~M.~Cline, K.~Kainulainen, P.~Scott and C.~Weniger,
  Phys.\ Rev.\ D {\bf 88}, 055025 (2013)
  Erratum: [Phys.\ Rev.\ D {\bf 92}, no. 3, 039906 (2015)]
  doi:10.1103/PhysRevD.92.039906, 10.1103/PhysRevD.88.055025
  [arXiv:1306.4710 [hep-ph]].
  
\bibitem{Alarcon:2011zs} 
  J.~M.~Alarcon, J.~Martin Camalich and J.~A.~Oller,
  Phys.\ Rev.\ D {\bf 85}, 051503 (2012)
  doi:10.1103/PhysRevD.85.051503
  [arXiv:1110.3797 [hep-ph]].

\bibitem{Alarcon:2012nr} 
  J.~M.~Alarcon, L.~S.~Geng, J.~Martin Camalich and J.~A.~Oller,
  Phys.\ Lett.\ B {\bf 730}, 342 (2014)
  doi:10.1016/j.physletb.2014.01.065
  [arXiv:1209.2870 [hep-ph]].
  
\bibitem{Aprile:2018dbl} 
  E.~Aprile {\it et al.} [XENON Collaboration],
  arXiv:1805.12562 [astro-ph.CO].

\bibitem{Ackermann:2015zua} 
  M.~Ackermann {\it et al.} [Fermi-LAT Collaboration],
  Phys.\ Rev.\ Lett.\  {\bf 115}, no. 23, 231301 (2015)
  doi:10.1103/PhysRevLett.115.231301
  [arXiv:1503.02641 [astro-ph.HE]].
  
\bibitem{AMS2t} M.~Aguilar {\it et al.} [AMS Collaboration],
Phys.\ Rev.\ Lett.\ {\bf 113}, 121102 (2014).
doi:10.1103/PhysRevLett.113.121102

\bibitem{AMS2ep} L.~Accardo {\it et al.} [AMS Collaboration],
Phys.\ Rev.\ Lett.\ {\bf 113}, 121101 (2014).
doi:10.1103/PhysRevLett.113.121101
  
\bibitem{Elor:2015bho} 
  G.~Elor, N.~L.~Rodd, T.~R.~Slatyer and W.~Xue,
  JCAP {\bf 1606}, no. 06, 024 (2016)
  doi:10.1088/1475-7516/2016/06/024
  [arXiv:1511.08787 [hep-ph]].
  
\bibitem{Khachatryan:2014rra} 
  V.~Khachatryan {\it et al.} [CMS Collaboration],
  Eur.\ Phys.\ J.\ C {\bf 75}, no. 5, 235 (2015)
  doi:10.1140/epjc/s10052-015-3451-4
  [arXiv:1408.3583 [hep-ex]].
  
  


\bibitem{Barducci:2016pcb} 
  D.~Barducci, G.~Belanger, J.~Bernon, F.~Boudjema, J.~Da Silva, S.~Kraml, U.~Laa and A.~Pukhov,
  Comput.\ Phys.\ Commun.\  {\bf 222}, 327 (2018)
  doi:10.1016/j.cpc.2017.08.028
  [arXiv:1606.03834 [hep-ph]].
  
\bibitem{CLs1} 
  A.~L.~Read, 
  J. Phys. {\bf G28}, (2002) 2693–2704.
  
\bibitem{CLs2}
  A. L. Read, “Modified frequentist analysis of search results (The CL(s) method),”
  in Workshop on confidence limits, CERN, Geneva, Switzerland, 17-18 Jan 2000:
  Proceedings. http://weblib.cern.ch/abstract?CERN-OPEN-2000-205.
  

  
\bibitem{Joyce:1994zt} 
  M.~Joyce, T.~Prokopec and N.~Turok,
  Phys.\ Rev.\ D {\bf 53}, 2958 (1996)
  doi:10.1103/PhysRevD.53.2958
  [hep-ph/9410282].
  
\bibitem{Cline:1997vk} 
  J.~M.~Cline, M.~Joyce and K.~Kainulainen,
  Phys.\ Lett.\ B {\bf 417}, 79 (1998)
  Erratum: [Phys.\ Lett.\ B {\bf 448}, 321 (1999)]
  doi:10.1016/S0370-2693(99)00033-7, 10.1016/S0370-2693(97)01361-0
  [hep-ph/9708393].
  
\bibitem{Cline:2000nw} 
  J.~M.~Cline, M.~Joyce and K.~Kainulainen,
  JHEP {\bf 0007}, 018 (2000)
  doi:10.1088/1126-6708/2000/07/018
  [hep-ph/0006119].
  
\bibitem{Fromme:2006wx} 
  L.~Fromme and S.~J.~Huber,
  JHEP {\bf 0703}, 049 (2007)
  doi:10.1088/1126-6708/2007/03/049
  [hep-ph/0604159].
  
\bibitem{NRecipe} 
  W.~H.~Press, S.~A.~Teukolsky, W.~T.~Vetterling and B.~P.~Flannery, 
  Numerical Recipes in C, 2nd edition, Cambridge University Press, 
  Cambridge, U.~K. (1992).
  
\bibitem{Cline:2011mm} 
  J.~M.~Cline, K.~Kainulainen and M.~Trott,
  JHEP {\bf 1111}, 089 (2011)
  doi:10.1007/JHEP11(2011)089
  [arXiv:1107.3559 [hep-ph]].

\bibitem{DOnofrio:2014rug} 
  M.~D'Onofrio, K.~Rummukainen and A.~Tranberg,
  Phys.\ Rev.\ Lett.\  {\bf 113}, no. 14, 141602 (2014)
  doi:10.1103/PhysRevLett.113.141602
  [arXiv:1404.3565 [hep-ph]].
  
  
\bibitem{Lew:1993yt} 
  H.~Lew and A.~Riotto,
  Phys.\ Lett.\ B {\bf 309}, 258 (1993)
  doi:10.1016/0370-2693(93)90930-G
  [hep-ph/9304203].
  
\bibitem{No:2011fi} 
  J.~M.~No,
  Phys.\ Rev.\ D {\bf 84}, 124025 (2011)
  doi:10.1103/PhysRevD.84.124025
  [arXiv:1103.2159 [hep-ph]].
  
\bibitem{Caprini:2015zlo} 
  C.~Caprini {\it et al.},
  JCAP {\bf 1604}, no. 04, 001 (2016)
  doi:10.1088/1475-7516/2016/04/001
  [arXiv:1512.06239 [astro-ph.CO]].

\bibitem{LISA} J.~Baker {\it et al.}, LISA science case document, Technical report, 2007.

\bibitem{BBO} E.~S.~Phinney {\it et al.}, NASA Mission Concept Study, Technical report, 2004.

\end{thebibliography}
\end{document}